\documentclass[sigconf]{acmart}
\usepackage{graphicx}
\usepackage{latexsym}
\usepackage{amsmath}
\usepackage{multirow}
\usepackage{array}
\usepackage{comment}
\usepackage{algorithm}
\usepackage{algpseudocode}
\usepackage[tikz]{mdframed}
\usepackage{anyfontsize}
\usepackage{sidecap}
\usepackage{rotating}
\usepackage{balance}
\usepackage{float}
\usepackage{subcaption}
\usepackage{colortbl}

\AtBeginDocument{%
  }

\setcopyright{acmlicensed}
\copyrightyear{2018}
\acmYear{2018}
\acmDOI{XXXXXXX.XXXXXXX}
\acmConference[Conference acronym 'XX]{Make sure to enter the correct
  conference title from your rights confirmation email}{June 03--05,
  2018}{Woodstock, NY}
\acmISBN{978-1-4503-XXXX-X/2018/06}

\begin{document}

\title{MemSifter: Offloading LLM Memory Retrieval via Outcome-Driven Proxy Reasoning}

\author{Jiejun Tan}
\email{zstanjj@ruc.edu.cn}
\orcid{0009-0001-8106-4780}

\author{Zhicheng Dou}
\authornote{Corresponding author.}
\email{dou@ruc.edu.cn}
\orcid{0000-0002-9781-948X}
\affiliation{%
  \department{Gaoling School of Artificial Intelligence}
\institution{Renmin University of China}
  \city{Beijing}
  \country{China}
}

\author{Liancheng Zhang}
\orcid{0009-0007-3017-2742}
\author{Yuyang Hu}
\orcid{0009-0007-2828-1165}
\author{Yiruo Cheng}
\orcid{0009-0004-0823-727X}
\affiliation{%
  \department{Gaoling School of Artificial Intelligence}
  \institution{Renmin University of China}
  \city{Beijing}
  \country{China}
}

\author{Ji-Rong Wen}
\email{jrwen@ruc.edu.cn}
\affiliation{%
  \department{Gaoling School of Artificial Intelligence}
\institution{Renmin University of China}
  \city{Beijing}
  \country{China}
}

\renewcommand{\shortauthors}{Jiejun Tan et al.}

\begin{abstract}
As Large Language Models~(LLMs) are increasingly used for long-duration tasks, maintaining effective long-term memory has become a critical challenge. Current methods often face a trade-off between cost and accuracy. Simple storage methods often fail to retrieve relevant information, while complex indexing methods (such as memory graphs) require heavy computation and can cause information loss. Furthermore, relying on the working LLM to process all memories is computationally expensive and slow.
To address these limitations, we propose MemSifter, a novel framework that offloads the memory retrieval process to a small-scale proxy model. Instead of increasing the burden on the primary working LLM, MemSifter uses a smaller model to reason about the task before retrieving the necessary information. This approach requires no heavy computation during the indexing phase and adds minimal overhead during inference.
To optimize the proxy model, we introduce a memory-specific Reinforcement Learning (RL) training paradigm. We design a task-outcome-oriented reward based on the working LLM's actual performance in completing the task. The reward measures the actual contribution of retrieved memories by mutiple interactions with the working LLM, and discriminates retrieved rankings by stepped decreasing contributions. Additionally, we employ training techniques such as Curriculum Learning and Model Merging to improve performance.
We evaluated MemSifter on eight LLM memory benchmarks, including Deep Research tasks. The results demonstrate that our method meets or exceeds the performance of existing state-of-the-art approaches in both retrieval accuracy and final task completion. MemSifter offers an efficient and scalable solution for long-term LLM memory. We have open-sourced the model weights, code, and training data to support further research.~\footnote{Codes are available at \url{https://github.com/plageon/MemSifter}}.
\end{abstract}



\received{20 February 2007}
\received[revised]{12 March 2009}
\received[accepted]{5 June 2009}

\maketitle

\section{Introduction}
With the rapid progress of Large Language Models~(LLMs), they show growing potential for long-duration tasks~\cite{hle,swe-bench,gaia}. However, effective memory over extended periods remains a key challenge~\cite{mem-survey}. As tasks lengthen, the accumulated interaction history soon exceeds the LLM's context window, forcing migration of information to persistent storage~\cite{memory-bank,longmemeval}. In this work, we focus on long-term persistent memory, rather than working memory. Working memory is transient, confined to the context window, and discarded at the end of a session~\cite{compress-context, memLLM}.

Following the taxonomy of \citet{mem-survey}, most existing work uses token-based forms to store LLM memory. The simplest, ``vanilla memory,'' uses a linear or flat memory bank that sequentially stores raw memory segments. At inference, the system retrieves the top-$k$ relevant pieces to concatenate into the context window. Vanilla memory is simple but faces major drawbacks: low retrieval accuracy and poor memory utilization. To address this, research has pursued two main directions: (1) \textbf{Structural Enhancement} during indexing, using extra computation to build richer indexes such as graphs~\cite{amem,h-mem,zep} or hierarchies; and (2) \textbf{Contextual Expansion} during inference, increasing the working LLM's input tokens.

Structural enhancement can improve retrieval diversity and accuracy over vanilla memory, but heavy indexing (summarization, entity extraction, and graph construction) adds substantial computation, and the abstraction process discards potentially important details. Moreover, because most stored memories are never reused, the upfront indexing cost is largely wasted.
Contextual expansion exploits the working LLM’s own capabilities. With precise awareness of task requirements, the LLM can read and use memories more accurately and efficiently than indexing-based methods, and its computation is demand-driven. However, since the working LLM is usually a heavy one, using it for both reading long memory contexts and task execution creates a dual burden.


To resolve this dilemma, we ask: \textit{Can we get the accuracy benefit of inference-time reasoning without adding more work to the primary LLM?} We propose \textbf{MemSifter}, a framework that \textbf{offloads} memory retrieval to a specialized, lightweight proxy model. The proxy uses ``reasoning-before-retrieval'': it reads the raw interaction history, analyzes the current task, and selects the critical evidence. The heavy working LLM then receives only the top-$k$ memory segments. By avoiding complex indexes and moving retrieval reasoning away from the main generation process, MemSifter improves recall with small overhead for the working LLM.

To bridge the gap between retrieval and task outcome, we propose a new Reinforcement Learning (RL) paradigm for training MemSifter. Unlike ``Think-then-Rerank'' methods~\cite{think-before-retrive,rearank,reasonrank}, which optimize isolated retrieval metrics using static relevance labels, we directly align memory proxy optimization with the working LLM's task success. This outcome-oriented design reflects two insights: (1) \textbf{Goal Alignment}: As an auxiliary module, the memory proxy should be judged only by its marginal contribution to downstream performance, not by proxy metrics like Recall or Precision. (2) \textbf{Label Scarcity}: Obtaining fine-grained gold rankings (e.g., exact top-20 orderings) for complex reasoning tasks is impractical, making supervised learning ineffective due to sparse signals.
However, training a retriever only on the working LLM's final task score has two challenges. First, it suffers from \textbf{ambiguous credit assignment}: a simple success signal cannot tell whether the correct answer came from retrieved memory or the LLM's internal knowledge. Second, a sparse scalar reward fails to capture \textbf{rank sensitivity}. Earlier ranks should receive more credit because they decide which evidence enters the working LLM's context first.

To address these issues, we design a reward mechanism comprising two components. We introduce \textbf{Marginal Utility Reward}, which isolates the genuine utility of retrieval by measuring the performance lift relative to a ``no-memory'' baseline, ensuring the retriever is rewarded only for bridging actual knowledge gaps. Furthermore, to enforce ranking quality without ground-truth labels, we employ a \textbf{Rank-Sensitive Reward}. By evaluating the LLM on varying top-$k$ cutoffs and applying diminishing weights (derived from DCG~\cite{dcg}), we incentivize the proxy model to prioritize critical evidence at the very top. Finally, to mitigate the instability inherent in RL, we employ an iterative training strategy. We incorporate warm-up supervision to address the cold-start problem, and utilize dynamic curriculum sampling alongside model merging to ensure diverse exploration and consistent performance gains.

We evaluate MemSifter on eight LLM memory datasets, spanning personal memory to long-horizon research tasks. MemSifter consistently surpasses prior state-of-the-art in both memory retrieval accuracy and end-to-end task performance. Our contributions are:
(1) MemSifter, a framework that offloads memory reasoning to a lightweight proxy, enabling efficient inference-time scaling without overloading the working LLM.
(2) A task-outcome-oriented RL paradigm that directly optimizes the proxy for the working LLM’s final success, tightly coupling retrieval and reasoning.
(3) State-of-the-art results on eight diverse benchmarks with superior efficiency, along with released code and data to support future work.

\section{Related Work}
\begin{figure*}
    \centering
    \includegraphics[width=\linewidth]{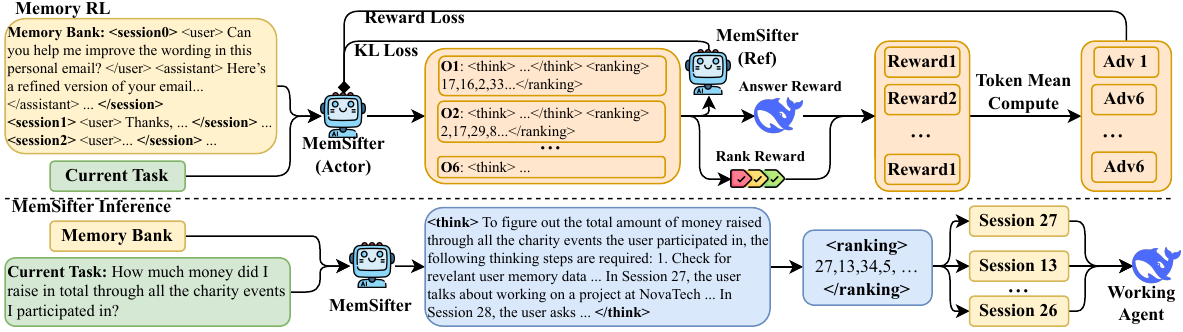}
    \caption{Top: the RL algorithm for MemSifter. Bottom: the inference pipeline for MemSifter.}
    \Description{A two-part diagram showing MemSifter training with task-outcome rewards and inference with a lightweight memory proxy before the working LLM.}
    \label{fig:refiner}
\end{figure*}

\subsection{Memory Mechanisms for LLMs}
Prior LLM memory research mainly studies architectures for persistent interaction, typically grouped into token-level, parametric, and latent memory.
\textbf{Token-level memory} stores information as external units such as text chunks or embeddings. Systems like
MemGPT~\cite{MemGPT} use hierarchical paging to cope with limited context windows, while Mem0~\cite{mem0} and others
standardize memory CRUD operations~\cite{Nemori, memory-bank}. Graph-based methods like GraphRAG~\cite{graphrag} and
HippoRAG~\cite{HippoRAG} build knowledge graphs for long-horizon, multi-hop reasoning. These approaches provide
explicit, editable storage but often incur high indexing costs and adapt poorly to dynamic, task-specific
retrieval~\cite{memmap, memory-matters}.
Recent RL-based memory agents is a special branch of token-level memory~\cite{memagent, memalpha}. They train an
agent or a working LLM to build and use memory across conversations.
They focus more on agent self-motivated long context folding and management. In comparison, MemSifter keeps
raw memory outside the working LLM and trains a separate proxy for retrieval offloading.
\textbf{Parametric memory} encodes knowledge in model parameters. Internal methods
like Character-LLM~\cite{CharacterLLM} fine-tune models with personas or behavioral priors, while external methods
like MLP-Memory~\cite{Wei2025MLPMA} inject retrieved knowledge into Transformer decoders via projection layers.
They offer fast, implicit access but updating memory requires
(re)training, limiting real-time use.
\textbf{Latent memory} compresses information into intermediate representations. RazorAttention~\cite{RazorAttention}
preserves effective attention spans, and MemoRAG~\cite{MemoRAG} creates compact global memory tokens. These are
efficient but often depend on complex architectural changes or lossy compression that can harm fine-grained
retrieval~\cite{memoryllm}.
MemSifter uses a hybrid design that combines token-level and parametric memory. It keeps raw content in a flexible
token-level store for real-time updates, while delegating memory processing and retrieval to the parametric weights
of a proxy model. Unlike purely parametric memory that stores facts, this proxy stores the
skill of identifying and reasoning over critical information in external storage.

\subsection{Inference-Time Retrieval Scaling}
To address the semantic limits of static embedding-based retrieval, recent work turns to
adding computation at query time to improve retrieval via reasoning and generation.
One direction is \textbf{reasoning-before-retrieval}. Standard encoders struggle with long-range dependencies and
complex implications. Methods like RGE~\cite{Liu2025ReasoningGE} and TTE~\cite{TTE} use Multimodal LLMs to generate
structured rationales or explanatory contexts before embedding. The O1 Embedder~\cite{O1Embedder} similarly inserts
a latent ``thinking'' phase, conditioning dense representations on explicit reasoning traces. These approaches trade
extra inference time for higher semantic fidelity, treating retrieval as a reasoning-centric process.
In parallel, \textbf{generative reranking} has become a key paradigm for final retrieval. Instead of vector similarity,
these methods use LLM reasoning to directly produce relevance rankings. Techniques span pointwise
scoring~\cite{HolisticEvaluationLMs} to listwise methods that reason over candidate
sets~\cite{EfficientReranking, RankPrompt, FoundintheMiddle}. Some rely on frozen LLMs with advanced prompting
~\cite{RankPrompt, AreOptimalAlg}), while others fine-tune models for
ranking via distillation or reinforcement learning~\cite{RankT5, RankR1, DistiLLM}.

Although effective, these approaches are computationally heavy, often invoking the same large models used for the
main task. MemSifter takes inspiration from both lines of work: it adopts the ``reasoning'' of inference-time
scaling and the ``listwise evaluation'' of generative reranking, but distills them into a specialized, lightweight
proxy. This preserves the benefits of reasoning-enhanced retrieval with minimal overhead.

\subsection{End-to-End Retrieval Optimization}
A key challenge in retrieval-augmented systems is the mismatch between pre-training objectives of retrieval (e.g.,
semantic similarity) and the downstream needs of generation~\cite{realm}. To address this, recent work optimizes
retrievers using feedback from the generator instead of static relevance labels.
Most methods derive supervision from the generator’s output to fine-tune retrieval, either by maximizing the
likelihood of ground-truth answers~\cite{replug, scarlet, atlas} or distilling knowledge from strong cross-encoder
rerankers~\cite{stochastic-rag, simlm}. More recent feedback-driven approaches use LLMs to annotate relevance or
create ``silver'' data, aligning retrieval scores with generation preferences~\cite{llm-specific-utility}.
These paradigms are limited for LLM memory. Standard RAG retrieves from static corpora for factual QA, whereas
autonomous LLMs must retrieve from their own episodic interaction history, which is dynamic, unstructured,
and context-dependent~\cite{mem1, memLLM}. Here, relevance is defined by utility for future planning, not factual
matching.
LLM memory tasks involve long-horizon, multi-step reasoning with no intermediate ground truths;
the only reliable signal is the sparse final outcome, making fine-grained supervision infeasible.

MemSifter instead uses a task-outcome-oriented RL paradigm. Rather than relying on intermediate
relevance labels or answer likelihoods, we optimize the memory proxy directly from the working LLM’s final success,
capturing the true “reasoning utility” of memories in the LLM’s context without human annotation or heavy computation.

\section{Methodology}


\begin{figure*}[t]
\begin{mdframed}[backgroundcolor=gray!5, roundcorner=3pt, innerleftmargin=10pt, innerrightmargin=10pt, innertopmargin=5pt, innerbottommargin=5pt, nobreak=true]

\small
\sloppy
\raggedright

\textbf{Historical Interaction Information:} <session0>...</session>...<session1>...

\textbf{Current Chat Context:} I'm trying to track my online spending, can you help me set up a budgeting tool or something?

You are tasked with a relevance-ranking task. Your core goal is to analyze two key inputs—the user's historical interaction information organized by sessions and the current chat context between the user and the AI assistant—then output the top 10 most relevant sessions ranked from highest to lowest relevance.
Please follow these strict guidelines to complete the task:

\textbf{1. Input Definitions} Historical Interaction Information: A collection of discrete, time-stamped sessions, each containing multiple interaction segments between the user and the AI (or other systems). Each session is identified by a session tag (e.g., <session1>, <session2>).
Current Chat Context: The ongoing real-time conversation between the user and the AI assistant, including all recent messages (user's latest questions/statements + the AI's latest responses, if any). This context reflects the user's current needs, concerns, or topics of interest.

\textbf{2. Relevance Judgment Criteria} To determine which sessions are most relevant, focus on the following priority factors (in descending order of importance):
Topic Consistency: Whether the core topic(s) of the historical session align with the user's current chat topic.
User Need Continuity: Whether the historical session reflects recurring or unresolved user needs related to the current chat.
Detail Overlap: Whether key entities, terms, or scenarios in the historical session match those in the current context.
Session Completeness: How well the entire session provides context and information relevant to the current query.

\textbf{3. Output Requirements} Format: Output the session identifiers of the top 10 most relevant sessions in descending order of relevance, session ids separated by comma, and wrap them with tag <ranking></ranking>. Ensure the final output contains exactly the session identifiers from the provided ``Historical Interaction Information''—do not create or modify session identifiers.
Accuracy: You should think carefully about the relevance of each session based on the above criteria before ranking them. Wrap your thinking process in ``<think>...</think>''.
Quantity: Output exactly 10 sessions if available; if fewer than 10 sessions exist, output all available sessions in ranked order.

\textbf{Brief Repeat:} During your reasoning process, do not repeat lengthy parts of the original message. If the original message is too long, you can omit some words in the middle by replacing them with ellipsis (...).
Example Output Format: <think>Your thinking process</think><ranking>27,13,34,5,12,8,21,45,6,19</ranking>
Please analyze the above two inputs and output the top 10 most relevant sessions in descending order of relevance, formatted as specified.

\end{mdframed}
\caption{The prompt for the memory proxy model, describing the task in detail.}
\Description{A framed text prompt that asks the proxy model to reason over historical sessions and output ranked session identifiers.}
\label{fig:ranking-prompt}
\end{figure*}

\subsection{Problem Formulation}
\label{problem-formulation}

To facilitate a precise formulation, we define the LLM memory task as follows. Let $M$ denote a working LLM, capable of interacting with human users and utilizing external tools. The memory bank stores the raw interaction history $H$ in persistent external disk, which captures the information exchange between the LLM and the external environment—ranging from user replies in assistant scenarios to tool execution outputs in research tasks. Formally, we represent the history as a sequence of sessions $H = \{s_1, \dots, s_N\}$. Each session $s_i$ is further composed of a sequence of interaction turns $s_i = \{t_1, \dots, t_{M_i}\}$.

Let $q$ denote the current task, representing the latest external input. The objective of the working LLM is to generate an optimal response $a$. Crucially, the quality of $a$ hinges upon the effective utilization of the long interaction history $H$. However, in long-horizon scenarios, processing the full history $H$ is often computationally prohibitive or exceeds the LLM's context window. To address this, we introduce a memory proxy $P$ to retrieve a subset of memory segments $M_\text{rel}$ relevant to $q$ from $H$. The working LLM then generates the final response $a = M(q, M_\text{rel})$ conditioned on both the current task $q$ and the retrieved memory segments.

\subsection{Memory Proxy Reasoning}
\label{memory-proxy-reasoning}



MemSifter operates on the interaction history $H$, which is initially stored in external storage. To facilitate retrieval, we first segment $H$ into distinct sessions based on topic continuity, where each session represents a coherent interaction sequence regarding a specific topic or tool usage. When a benchmark provides session boundaries, we keep them; for deep research trajectories, we treat each search or reasoning step as one session. We reformat these sessions by wrapping them with unique identifiers, e.g., ``\texttt{<session~$i$> ... </session>}'', enabling the proxy model to distinguish between different historical contexts.

During the loading phase, if the total token count of $H$ exceeds the proxy model's context window (e.g., 128k tokens),
we employ a lightweight embedding model to perform a \textbf{coarse-grained pre-filtering}. This step ranks
sessions by query-session cosine similarity and keeps the top sessions under the token budget, without a fixed
similarity threshold. More filtering details are in Appendix~\ref{sec:appendix-experimental-details}.

The memory proxy then performs a ``Think-and-Rank'' inference process on this loaded context. Given the current task $q$ and the formatted history, it first generates a reasoning rationale $t$ (enclosed in ``\texttt{<think>}'' tags) to analyze dependencies. It then predicts the IDs of the top-$k$ most relevant sessions, enclosed in ``\texttt{<ranking>}'' tags (see Figure~\ref{fig:ranking-prompt} for the prompt). We use a fixed top-10 ranking list for fair comparison, and all methods follow the same top-$k$ budget. Finally, the content of these top-$k$ sessions is retrieved and concatenated with $q$ to form the context for the working LLM, which then generates the final response $a$.

\begin{figure*}[t]
    \centering
    \begin{subfigure}[b]{0.5\textwidth}
        \centering
        \includegraphics[width=\linewidth]{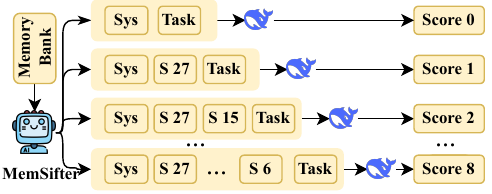}
        \caption{Marginal Utility Reward}
        \label{fig:reward}
    \end{subfigure}%
    \hfill
    \begin{subfigure}[b]{0.43\textwidth}
        \centering
        \includegraphics[height=3.8cm]{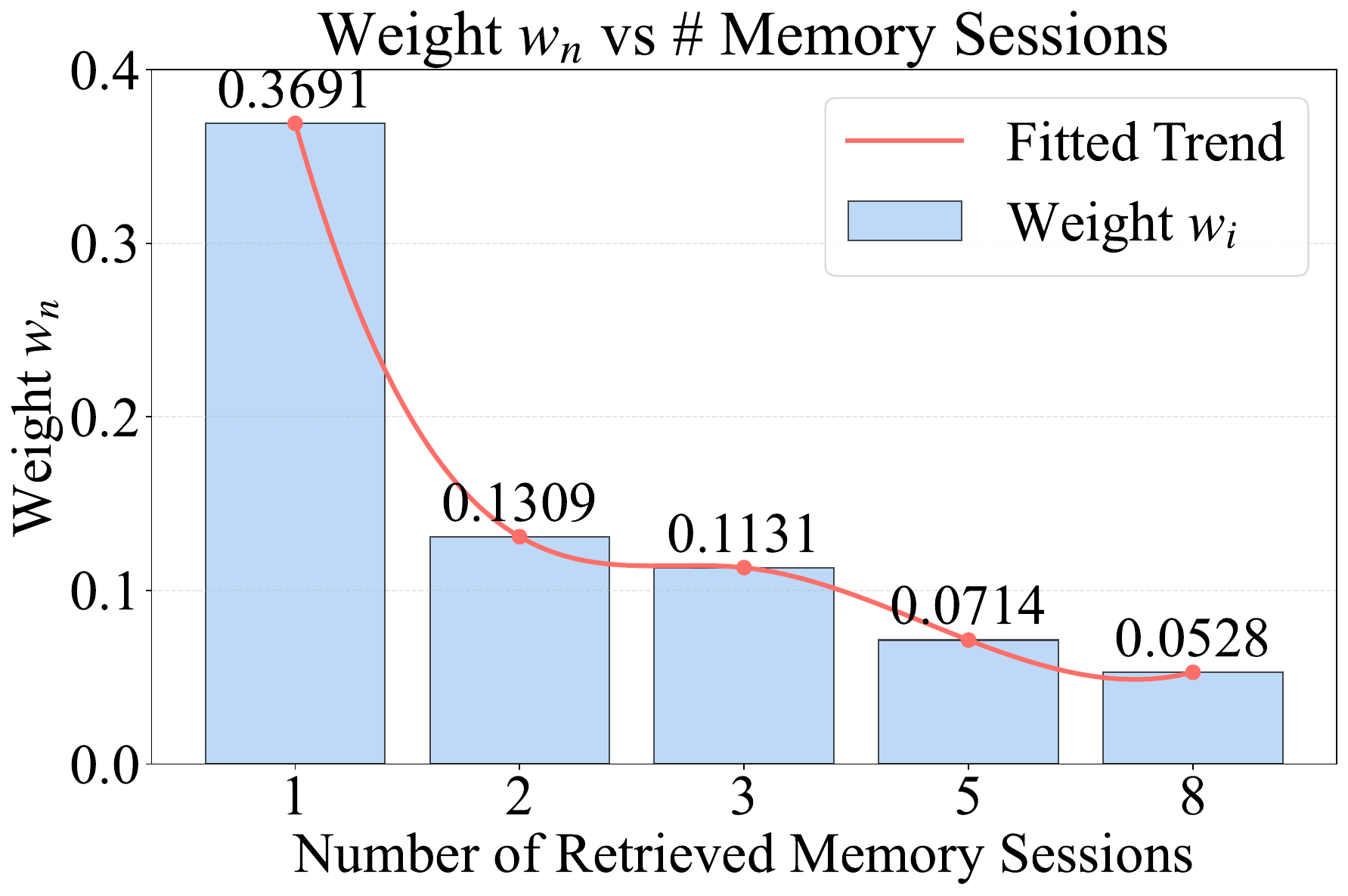}
        \caption{Diminishing Weight $w_n$}
        \label{fig:fib-graph}
    \end{subfigure}%

    \caption{Task-oriented reward design.}
    \Description{Two panels illustrating the marginal utility reward computation and the diminishing rank weights used in the task-oriented reward.}
    \label{fig:three_images}
\end{figure*}


\subsection{Task-Outcome-Oriented Reward}
\label{task-outcome-oriented-reward}

Retrieval-metric-based rewards work well in ``Think-and-Rerank'' settings but do not transfer directly to our framework. Our memory proxy’s main goal is to enhance the working LLM, so retrieval quality should be judged by its impact on final task performance, not by intrinsic retrieval metrics. Fine-grained supervision is also difficult: most tasks lack ground-truth memory annotations, and available annotations rarely provide full top‑$k$ rankings as in reranking datasets. To overcome these issues, we derive a reward signal from the working LLM’s final performance, using two designs: utility estimation via ablation and rank-sensitive rewards.





\subsubsection{\textbf{Marginal Utility Reward}}
\label{ablation-based-marginal-utility}

The first design component aims to quantify the genuine utility of retrieved memory. We adopt the principle of ablation
study to isolate the net contribution of memory segments. Specifically, we define $s_0$ as the baseline performance
score where the working LLM attempts the task \textit{without} any retrieved memory. When the working LLM is provided
with the top-$k$ retrieved segments, the score is denoted as $s_k$. The true effectiveness of the memory retrieval is thus quantified by the performance lift $s_k - s_0$. To further capture the granularity of retrieval quality, we extend this global ablation to a progressive evaluation strategy. Instead of a single static top-$k$ check, we evaluate the LLM's performance across a sequence of increasing context sizes, denoted as $\mathcal{K} = \{k_1, k_2, \dots, k_N\}$, as shown in Figure~\ref{fig:reward}.

Evaluating the working LLM at every possible rank position would incur prohibitive computational costs. To mitigate this, we employ two optimization strategies. First, we adopt a \textbf{Fibonacci sampling sequence} (e.g., $\{1, 2, 3, 5, \dots\}$) for $\mathcal{K}$. This reduces inference calls from linear to logarithmic in list size while still capturing key performance inflection points at the top of the list. Second, during training, we use an efficient but capable working LLM~(Qwen3-30B-A3B), greatly speeding up the feedback loop.

For any two consecutive evaluation tiers $k_{n-1}$ and $k_n$, the performance difference $\Delta s_n = s_{k_n} - s_{k_{n-1}}$ explicitly represents the marginal contribution provided by the newly added memory segments ranked between $k_{n-1}$ and $k_n$. 
A potential concern is that this incremental calculation might unfairly penalize lower-ranked memories that depend on earlier information. However, in memory-based generation, utility is inherently context-dependent: a memory’s value lies in how it complements preceding context to complete a reasoning chain. By computing the marginal lift $\Delta s_n$ given the top-$k_{n-1}$ memories, our reward mechanism pushes the retriever to place foundational information first (to secure an early score $s_{k_1}$) and complementary details next, thereby building a coherent, logically progressive context for the working LLM.

\subsubsection{\textbf{Rank-Sensitive Reward}}
\label{rank-sensitive-reward-shaping}
The progressive ablation above measures how much information is gained. It treats all gains equally regardless
of their position. For retrieval, earlier ranks should receive more credit because they define which evidence is seen
first by the working LLM. This reward shapes retrieval priority. It is not a claim about the working LLM's full
attention pattern. We use a \textbf{diminishing returns} design, analogous to the DCG metric~\cite{dcg}.

We posit that the reward for a marginal performance gain $\Delta s_n$ should be discounted based on its rank position. In other words, a performance boost achieved by the top-1 memory is rewarded heavily, while the same boost achieved by the top-5 memory receives less credit. Mathematically, we formulate the final reward $R_\text{ans}$ as a weighted aggregation of the progressive scores:
\begin{equation}
\label{eq:final_reward}
R_\text{ans} = \sum_{n=1}^N \gamma_n \cdot (s_{k_n} - s_{k_{n-1}}).
\end{equation}

Following the same intuition, we use the DCG-style form:
\begin{equation}
    R_\text{ans} = \sum_{i=1}^K \frac{c_i}{\log_2 (i+1)},
\end{equation}
where $\gamma_n$ is a decaying weight coefficient ensuring that early ranks dominate the reward signal. By rearranging the terms and normalizing against the baseline $s_0$, this can be reformulated into a computationally efficient form:
\begin{equation}
R_\text{ans} = -s_0 + \sum_{n=1}^N w_n \cdot s_{k_n},
\end{equation}
where the coefficients $w_n$ follow the logarithmic decay of DCG (i.e., $\propto 1/\log_2(rank)$). This design rewards the proxy for putting useful evidence near the top of the list.


Ideally, the standard DCG formulation scales the contribution of each individual document at rank $i$ by a logarithmic decay factor $1 / \log_2(i+1)$. However, as described in Sec.~\ref{ablation-based-marginal-utility}, we employ a sparse Fibonacci sampling strategy for efficiency. Consequently, each evaluation step $n$ introduces a \textit{batch} of new memory sessions (ranging from rank $k_{n-1}+1$ to $k_n$) rather than a single entry. Since we only observe the aggregate performance lift $\Delta s_n$ for this entire batch, we approximate the DCG summation by applying the decay factor of the batch's outer boundary $k_n$ to the collective gain:
\begin{equation*}
\sum_{i=k_{n-1}+1}^{k_n} \frac{c_i}{\log_2 (i+1)} \approx \frac{1}{\log_2 (k_n+1)} \cdot \sum_{i=k_{n-1}+1}^{k_n} c_i = \frac{s_{k_n} - s_{k_{n-1}}}{\log_2 (k_n+1)}.
\end{equation*}

To formalize this, we approximate the theoretical DCG reward by assigning a unified discount factor to each evaluation batch. Let $D_n = \frac{1}{\log_2 (k_n + 1)}$ represent the discount factor for the memory segments up to rank $k_n$. We approximate the total reward as the sum of discounted marginal gains:
\begin{equation}
\label{eq:approx_reward}
R_\text{ans} \approx \sum_{n=1}^N D_n \cdot (s_{k_n} - s_{k_{n-1}}),
\end{equation}
where we define the boundary conditions $k_0=0$ and $s_{k_0}=s_0$.

Directly computing Equation~(\ref{eq:approx_reward}) requires calculating differences step-by-step. To simplify this for efficient computation, we regroup the terms by their performance scores $s_{k_n}$. This transformation isolates the baseline term $s_0$, which carries a coefficient of $-D_1$. Given our setting of $k_1=1$, we have $D_1 = \frac{1}{\log_2 2} = 1$, effectively reducing the baseline term to a simple subtraction of $s_0$.

Consequently, the equation can be rewritten as a weighted sum of the accumulated scores, $R_\text{ans} = -s_0 + \sum_{n=1}^N w_n s_{k_n}$, where the weight $w_n$ represents the \textit{differential discount} between consecutive tiers, as visualized in Figure~\ref{fig:fib-graph}. Specifically, $w_n$ is defined as:
\begin{equation}
w_n =
\begin{cases}
\dfrac{1}{\log_2 (k_n + 1)} - \dfrac{1}{\log_2 (k_{n+1} + 1)}, & \text{if } 1 \le n < N ; \\
\dfrac{1}{\log_2 (k_n + 1)}, & \text{if } n = N .
\end{cases}
\end{equation}

Substituting these weights back into Equation~(\ref{eq:final_reward}) gives larger weights to gains at earlier ranks (smaller $k_n$) and smaller weights to gains at later ranks. This aligns our reward mechanism with DCG's diminishing returns property and encourages the memory proxy to place critical information early.

\subsection{Optimization Protocol and Curriculum}
\label{sec:training_strategy}

We employ an iterative training strategy to progressively enhance the retriever's capabilities.

\textbf{Dynamic Curriculum Construction}
To maximize sample efficiency, we construct a dynamic training set before each iteration. We define an anchor score $\tau$ (empirically set to $0.2$) to identify tasks within the model's \textbf{``zone of proximal development''}—specifically, difficult tasks where the model exhibits emerging capabilities but has not yet achieved mastery. We prioritize selecting task training samples where the model's current performance aligns most closely with $\tau$. Furthermore, during the RL rollout phase, we incorporate a dynamic sampling strategy inspired by DAPO~\cite{dapo} to ensure the generated trajectories maintain appropriate difficulty levels, preventing the model from overfitting to easy samples or collapsing on impossible ones.

\textbf{Hybrid Reward with Cold Start}
Although our ultimate goal is to optimize the task-outcome-oriented score $R_\text{ans}$, an RL based solely on outcomes can suffer instability during the early stages of training (the cold-start problem). To mitigate this and accelerate convergence, we introduce a temporary auxiliary objective.
During initial warm-up steps, we utilize a small set of available memory ranking annotations to compute a retrieval quality metric, $R_{ret}$ (measured by DCG). This phase is strictly designed to help the proxy model learn the correct output format and basic semantic relevance. The total reward $R$ is formulated as a weighted sum: $R = \alpha \cdot R_\text{ans} + \beta \cdot R_{ret},$
where $\alpha$ and $\beta$ are hyperparameters. To maintain our label-free design, we use a \textbf{reward annealing schedule}: $\beta$ is high only in the first epoch and then annealed to zero, shifting the model to purely outcome-oriented optimization on $R_\text{ans}$.

\textbf{Stabilization via Model Averaging}
To reduce RL training fluctuations, we apply \textbf{Model Averaging} at the end of each iteration. We select the top-$k$ checkpoints by validation performance and compute the arithmetic mean of their trainable parameters. The merged model initializes the next iteration, consolidating learned capabilities and smoothing optimization.

\section{Experiments}

\begin{table*}[t]
\small
\centering
\caption{Main Results of MemSifter evaluated by F1 score. The best and second best of each model are in \textbf{bold} and \underline{underlined}.}
\label{tab:main-results}
{
\begin{tabular}{llcccccccccc}
\toprule
& \multirow{2}{*}{\textbf{Method}} & \multicolumn{1}{c}{\textbf{LoCoMo}} & \multicolumn{2}{c}{\textbf{LongMemEval}} & \multicolumn{2}{c}{\textbf{PersonaMem}} & \multicolumn{1}{c}{\textbf{PerM-V2}} & \multicolumn{1}{c}{\textbf{ZH4O}} & \multicolumn{1}{c}{\textbf{HotpotQA}} & \multicolumn{1}{c}{\textbf{WebWalker}} & \multicolumn{1}{c}{\textbf{WebDancer}} \\
\cmidrule(lr){3-3}\cmidrule(lr){4-5}\cmidrule(lr){6-7}\cmidrule(lr){8-8}\cmidrule(lr){9-9}\cmidrule(lr){10-10}\cmidrule(lr){11-11}\cmidrule(lr){12-12}
& & 32K & 128K & 1M & 32K & 128K & 128K & 128K & 128K & 128K & 128K \\
\midrule
\multirow{12}{*}{\rotatebox{90}{\textbf{DeepSeek V3.2}}} 
& BGE-M3 & 29.49 & 31.62 & 29.91 & 20.76 & 18.83 & 21.45 & 45.38 & 20.79 & 22.61 & 33.89 \\
& GemmaEmb & 29.90 & 32.62 & 30.83 & 20.96 & 18.22 & 22.95 & 41.84 & 21.01 & 21.69 & \underline{37.82} \\
& Nemori & 27.43 & 34.39 & 29.24 & 21.25 & 19.89 & \textbf{23.72} & 45.07 & 22.76 & 24.39 & 35.28 \\
& Mem0 & 31.62 & 33.93 & 28.20 & 22.33 & 20.63 & 21.63 & 47.19 & 20.35 & 23.46 & 32.58 \\
& MemAgent & 37.36 & \underline{35.16} & 31.60 & 20.62 & 19.84 & 21.74 & 41.15 & \underline{24.09} & 24.05 & 34.10 \\
& Mem-$\alpha$ & \underline{38.70} & 34.71 & \underline{33.07} & \underline{23.37} & 20.27 & 21.01 & \underline{47.27} & 23.02 & \underline{24.49} & 34.58 \\
& HippoRAG & 28.11 & 33.64 & 30.83 & 22.65 & 19.24 & 22.94 & 46.56 & 23.20 & 24.19 & 33.26 \\
& A-MEM & 31.82 & 31.96 & 32.49 & 21.44 & 18.14 & 23.04 & 45.69 & 23.12 & 21.88 & 36.72 \\
& Rearank & 33.04 & 33.53 & 32.47 & 22.23 & 18.41 & 19.76 & 47.14 & 19.76 & 24.19 & 37.23 \\
& ReasonRank & 31.94 & 34.31 & 28.93 & 21.60 & \underline{20.74} & 22.27 & 45.55 & 22.25 & 22.79 & 32.95 \\
& DS-V3.2 & 30.97 & 35.15 & 30.89 & 21.69 & 20.15 & 21.55 & 46.73 & 19.88 & 22.49 & 34.31 \\
\rowcolor[RGB]{235,245,250} & MemSifter & \textbf{41.79} & \textbf{35.38} & \textbf{33.32} & \textbf{23.70} & \textbf{21.14} & \underline{23.57} & \textbf{48.13} & \textbf{24.95} & \textbf{26.11} & \textbf{38.21} \\
\midrule
\multirow{12}{*}{\rotatebox{90}{\textbf{Qwen3-30B-A3B-Ins}}} 

& BGE-M3 & 38.73 & 41.75 & \underline{49.07} & 24.98 & 22.73 & 20.21 & 48.31 & 20.49 & 25.01 & 31.85 \\
& GemmaEmb & 36.25 & 43.04 & 48.16 & 23.30 & 22.36 & 20.12 & 50.25 & 19.84 & 24.51 & 32.45 \\
& Nemori & 40.96 & \underline{46.09} & 47.31 & 24.35 & 22.69 & 18.62 & 45.53 & 20.91 & 25.80 & 32.77 \\
& Mem0 & 36.20 & 38.56 & 48.53 & 24.87 & 22.51 & 21.11 & 48.40 & 20.57 & 24.24 & 32.88 \\
& MemAgent & \underline{42.90} & 45.95 & 48.25 & \underline{26.08} & 22.35 & 20.47 & \textbf{52.45} & 19.80 & 27.04 & 32.25 \\
& Mem-$\alpha$ & 40.63 & 45.86 & 44.45 & 24.04 & 22.66 & 22.72 & 48.11 & 22.08 & \underline{27.06} & 30.74 \\
& HippoRAG & 41.94 & 40.13 & 47.62 & 23.35 & 21.31 & 20.62 & 48.36 & 21.41 & 25.16 & 31.79 \\
& A-MEM & 35.87 & 40.46 & 46.77 & 24.57 & 20.78 & 21.07 & 44.52 & 19.85 & 26.13 & \underline{33.67} \\
& Rearank & 39.19 & 39.55 & 45.60 & 22.37 & \underline{22.95} & 22.17 & 49.02 & 19.95 & 24.14 & 31.40 \\
& ReasonRank & 41.04 & 44.26 & 48.20 & 23.78 & 20.18 & 19.05 & 47.16 & \underline{22.37} & 25.72 & 30.00 \\
& Qwen3-30B & 39.81 & 42.20 & 46.42 & 24.25 & 20.82 & \textbf{24.49} & \underline{51.68} & 21.59 & 25.25 & 33.63 \\
\rowcolor[RGB]{235,245,250}& MemSifter & \textbf{46.39} & \textbf{47.26} & \textbf{49.58} & \textbf{26.45} & \textbf{23.75} & \underline{22.81} & 50.91 & \textbf{22.71} & \textbf{27.44} & \textbf{35.10} \\
\bottomrule
\end{tabular}
}
\end{table*}

\subsection{Datasets}

We select five representative personal LLM datasets~(1\textasciitilde5) and three deep research~(6\textasciitilde8) to evaluate our method and baselines: 

(1)~LoCoMo~\cite{locomo}: A benchmark for evaluating very long-term conversational memory, featuring 10 multimodal dialogues averaging 300 turns to assess understanding of facts and temporal relationships; 
(2)~LongMemEval~\cite{longmemeval}: A comprehensive benchmark designed to evaluate five core long-term memory capabilities in continuous chatbot interactions; 
(3)~PersonaMem~\cite{personamem}: A dataset with over 180 curated personas that assesses an LLM's ability to internalize user profiles from long interaction histories; 
(4)~PerM-V2~\cite{Jiang2025PersonaMemv2TP}: A scaled-up version simulating 1,000 user scenarios, focusing on enabling models to implicitly infer preferences from long conversations; 
(5)~ZH4O~\cite{PerLTQA}: A QA dataset integrating semantic and episodic memories, covering diverse long-term memory types in mixed-context scenarios; 
(6)~HotpotQA~\cite{hotpot-qa}: A dataset requiring multi-hop reasoning across multiple supporting documents to answer complex natural language questions; 
(7)~WebWalker~\cite{webwalker}: A benchmark assessing the capacity of agents to systematically traverse website subpages and extract multi-layered, high-quality information; 
(8)~WebDancer~\cite{webdancer}: A dataset focused on autonomous multi-step research, providing rich browsing trajectories for training agents in in-depth information seeking tasks.

In our experiments, we randomly sample 400 questions from the test set of the LoCoMo, PersonaMem, PersonaMem-v2, and PerLTQA datasets for evaluation. For LongMemEval, we randomly sample 150 questions for the test set.
Detailed train/test split statistics are provided in Appendix~\ref{sec:appendix-experimental-details}. We report end-to-end generation F1 as the unified metric across all benchmarks because every method ultimately produces a free-form answer from the retrieved memory context. For PersonaMem, we additionally use option-normalized multiple-choice accuracy; detailed accuracy results are reported in Appendix~\ref{sec:appendix-experimental-details}. Retrieval quality is evaluated with NDCG@1 and NDCG@5, and efficiency is measured by input/output token counts and latency.


\subsection{Baselines}

We compare MemSifter against state-of-the-art baselines across five distinct categories.
\textbf{Embedding-based retrieval:}
(1) BGE-M3~\cite{BGEM3}: A unified retrieval model supporting dense, sparse, and multi-vector representations, capable of handling long inputs up to 8,192 tokens.
(2) EmbeddingGemma~\cite{EmbeddingGemma}: A lightweight embedding model distilled from the Gemma family, optimizing semantic representation through geometric alignment for low-latency retrieval.
\textbf{LLM Memory Frameworks:}
(1) Mem0~\cite{mem0}: A long-term memory management layer that proactively extracts, stores, and updates salient information as discrete memory units.
(2) Nemori~\cite{Nemori}: A cognitively inspired architecture that segments interaction history into coherent units and dynamically updates them via a self-organizing mechanism.
(3) MemAgent~\cite{memagent}: An RL-based memory agent that trains the working LLM to maintain and exploit memory across multiple conversations.
(4) Mem-$\alpha$~\cite{memalpha}: An RL-based memory-construction agent that learns when and how to write structured memories from long interaction histories.
\textbf{Graph Retrieval:}
(1) HippoRAG~\cite{HippoRAG}: A neuro-symbolic retrieval framework that constructs knowledge graphs from documents and utilizes Personalized PageRank to enable multi-hop associative recall.
(2) A-MEM~\cite{amem}: An evolving memory system that structures summarized experiences into a dynamic graph to support reasoning over complex LLM histories.
\textbf{Generative Rerankers:}
(1) Rearank~\cite{rearank}: An RL-based listwise reranking LLM that explicitly reasons about candidate relevance to optimize the final ranking.
(2) ReasonRank~\cite{reasonrank}: A reasoning-enhanced reranker trained via supervised and reinforcement learning to generate explanation-aware relevance scores.
For fairness, Rearank and ReasonRank are retrained on the same training pool as MemSifter using their original single-round RL reward designs. MemAgent and Mem-$\alpha$ are adapted as RL-memory baselines under the same persistent-memory benchmark interface: each method constructs or maintains memory from the historical sessions, and the working LLM answers with the resulting memory context under the same top-$k$ budget.
\textbf{Long-context LLMs (Retrieval-Free):}
(1) Qwen3-30B~\cite{qwen3technicalreport}: A large-scale Mixture-of-Experts model leveraging a massive context window to directly process full interaction histories without external retrieval.
(2) DeepSeek-V3.2~\cite{DeepSeekV3.2}: A high-capacity model optimized for long-context efficiency, serving as a strong baseline for native in-context learning capabilities.

\subsection{Main Results}

To verify the effectiveness of MemSifter, we evaluate our method against the baseline methods on the eight aforementioned datasets. Table~\ref{tab:main-results} presents the main F1 results of the end-to-end experiments, with adapted RL-memory baselines included as explicit comparison rows. The proxy is trained with feedback from Qwen3-30B-A3B-Instruct and evaluated with both Qwen3-30B-A3B-Instruct and DeepSeek V3.2, which tests transfer across working LLMs. Based on the experimental results, we draw the following analysis.
(1)~\textbf{Superiority over Embedding:} Compared to standard embedding retrieval methods, MemSifter achieves significantly higher retrieval precision. By filtering out irrelevant noise that naive embeddings often capture, our method provides the working LLM with a cleaner context, directly translating to substantial gains in final task success rates.
(2)~\textbf{Task Utility vs. Generative Rerank:} MemSifter outperforms ``Think-then-Rerank'' baselines. This confirms a critical hypothesis: \textit{semantic relevance does not equal task utility}. While rerankers optimize for query-document similarity, our outcome-oriented reward ensures that the retrieved memory specifically aids in solving the reasoning problem at hand.
(3)~\textbf{Efficiency against Complex Architectures:} Unlike complex memory pipelines (e.g., graph-based or generative memory) that incur high latency and indexing overheads, MemSifter achieves state-of-the-art performance with a lightweight architecture. This demonstrates that a well-optimized, small-scale proxy can surpass heavy, complex systems in practical scenarios.
(4)~\textbf{Effective Alternative to Long-Context Models:} Compared to directly feeding full history into long-context LLMs, MemSifter reduces computational costs while maintaining or even exceeding performance. Long-context models often struggle with the ``lost-in-the-middle'' phenomenon; our method mitigates this by proactively selecting and reasoning about critical information before the working LLM processes it.

\subsection{Ablation Study}

To assess the effectiveness of our reward mechanism's core components, we run ablations by removing individual design elements.  
(1)~\textbf{w/o Task-Outcome Reward}: We remove RL-based optimization and train the proxy only on retrieval-metric supervision (as in warm-up). The large performance drop shows that optimizing static relevance (e.g., semantic similarity) is insufficient for downstream utility: without outcome-based signals, the proxy misses memories that are semantically distant but logically crucial for LLM reasoning.  
(2)~\textbf{w/o Marginal Utility}: We replace the incremental gain metric with the absolute task score from using top-$k$ memories, omitting the “no-memory’’ baseline. This hurts performance due to \textit{ambiguous credit assignment}: the proxy gets high rewards even on easy tasks solved by parametric knowledge alone, obscuring the true contribution of retrieved information and destabilizing training.  
(3)~\textbf{w/o Rank-Sensitive Weights}: We replace the DCG-weighted sum with a simple mean over top-$k$ evaluations. The resulting degradation shows the value of \textit{rank sensitivity}: uniform averaging ignores position and does not reward the proxy for placing key evidence early.
(4)~\textbf{w/o Thinking / Direct Rank}: We add a controlled direct-rank setting to isolate the role of explicit reasoning before retrieval. The result row is included in Table~\ref{tab:ablation-study}.

\begin{table*}[t]
\small
\centering
\caption{Main Memory Retrieval Results of MemSifter. The best and second best of each model are in \textbf{bold} and \underline{underlined}.}
\label{tab:retrieval-results}
\setlength{\tabcolsep}{1.8mm}{
\begin{tabular}{lcccccccccccc}
\toprule
\multirow{2}{*}{\textbf{Method}} & \multicolumn{2}{c}{\textbf{LoCoMo}} & \multicolumn{4}{c}{\textbf{LongMemEval}} & \multicolumn{2}{c}{\textbf{PersonaMem}} & \multicolumn{2}{c}{\textbf{PerM-V2}} & \multicolumn{2}{c}{\textbf{ZH4O}} \\
\cmidrule(lr){2-3}\cmidrule(lr){4-7}\cmidrule(lr){8-11}\cmidrule(lr){12-13}
& \multicolumn{2}{c}{32K} & \multicolumn{2}{c}{128K} & \multicolumn{2}{c}{1M} & \multicolumn{2}{c}{128K} & \multicolumn{2}{c}{128K} & \multicolumn{2}{c}{128K} \\ 
\midrule
    Metric &     ndcg@1 &     ndcg@5 &     ndcg@1 &     ndcg@5 &     ndcg@1 &     ndcg@5 &     ndcg@1 &     ndcg@5 &     ndcg@1 &     ndcg@5 &     ndcg@1 &     ndcg@5  \\
\midrule
BGE-M3 & 36.25 & 58.13 & 56.00 & 87.90 & 38.67 & 67.07 & 28.25 & 52.34 & 12.00 & 18.95 & 25.75 & 39.06 \\
GemmaEmb & 39.64 & \underline{70.76} & 60.51 & \underline{89.64} & \underline{50.83} & \underline{70.78} & \underline{37.07} & 54.72 & 12.18 & 19.42 & \underline{28.75} & 41.14 \\
ReaRank & 43.18 & 59.13 & \underline{62.70} & 88.50 & 39.33 & 68.19 & 33.86 & \underline{55.74} & \underline{13.68} & \textbf{20.76} & 25.43 & 40.37 \\
ReasonRank & \underline{47.64} & 64.94 & 60.81 & 88.18 & 45.14 & 69.32 & 30.59 & 55.18 & 11.26 & 19.94 & 24.49 & \underline{41.34} \\
\rowcolor[RGB]{235,245,250} MemSifter & \textbf{70.00} & \textbf{78.11} & \textbf{67.33} & \textbf{89.67} & \textbf{60.00} & \textbf{75.63} & \textbf{43.50} & \textbf{61.70} & \textbf{15.00} & \underline{20.09} & \textbf{40.50} & \textbf{50.97} \\
\bottomrule
\end{tabular}
}
\end{table*}

\begin{table*}[t]
\small
\centering
\caption{Ablation Study of MemSifter with DeepSeek V3.2. The best and second best of each dataset are in \textbf{bold} and \underline{underlined}.}
\label{tab:ablation-study}
\setlength{\tabcolsep}{0.5mm}{
\begin{tabular}{lcccccccc}
\toprule
\multirow{2}{*}{\textbf{Method}} & \multicolumn{1}{c}{\textbf{LoCoMo}} & \multicolumn{1}{c}{\textbf{LongMemEval}} & \multicolumn{1}{c}{\textbf{PersonaMem}} & \multicolumn{1}{c}{\textbf{PerM-V2}} & \multicolumn{1}{c}{\textbf{ZH4O}} & \multicolumn{1}{c}{\textbf{HotpotQA}} & \multicolumn{1}{c}{\textbf{WebDancer}} & \multicolumn{1}{c}{\textbf{WebWalker}} \\
\cmidrule(lr){2-2}\cmidrule(lr){3-3}\cmidrule(lr){4-4}\cmidrule(lr){5-5}\cmidrule(lr){6-6}\cmidrule(lr){7-7}\cmidrule(lr){8-8}\cmidrule(lr){9-9}
& 32K & 1M & 128K & 128K & 128K & 128K & 128K & 128K \\ 
\midrule
\rowcolor[RGB]{235,245,250} \textbf{MemSifter} & \textbf{41.79} & \textbf{33.32} & \textbf{21.14} & \textbf{23.57} & \textbf{48.13} & \textbf{24.95} & \textbf{26.11} & \textbf{38.21} \\
\quad w/o Outcome Reward & 30.59 (26.80\%$\downarrow$) & 28.67 (13.96\%$\downarrow$) & 15.27 (27.77\%$\downarrow$) & 18.26 (22.53\%$\downarrow$) & 40.05 (16.79\%$\downarrow$) & 21.33 (14.51\%$\downarrow$) & 20.03 (23.29\%$\downarrow$) & 30.16 (21.07\%$\downarrow$) \\
\quad w/o Marginal Utility & 38.62 (7.59\%$\downarrow$) & 32.67 (1.95\%$\downarrow$) & 18.66 (11.73\%$\downarrow$) & 21.03 (10.78\%$\downarrow$) & 46.74 (2.89\%$\downarrow$) & 23.60 (5.41\%$\downarrow$) & 24.70 (5.40\%$\downarrow$) & 35.96 (5.89\%$\downarrow$) \\
\quad w/o Reward Shaping & \underline{40.68} (2.66\%$\downarrow$) & \underline{33.26} (0.18\%$\downarrow$) & \underline{19.89} (5.91\%$\downarrow$) & \underline{22.42} (4.88\%$\downarrow$) & \underline{47.30} (1.72\%$\downarrow$) & \underline{24.87} (0.32\%$\downarrow$) & \underline{25.12} (3.79\%$\downarrow$) & \underline{37.53} (1.78\%$\downarrow$) \\
\quad w/o Thinking & 35.34 (15.43\%$\downarrow$) & 29.81 (10.53\%$\downarrow$) & 16.14 (23.65\%$\downarrow$) & 19.07 (19.09\%$\downarrow$) & 43.36 (9.91\%$\downarrow$) & 20.69 (17.07\%$\downarrow$) & 21.31 (18.38\%$\downarrow$) & 32.28 (15.52\%$\downarrow$) \\

\bottomrule
\end{tabular}
}
\end{table*}

\subsection{Further Analysis}

\subsubsection{\textbf{Retrieval Quality Analysis}}

To validate that our end-to-end gains stem from superior context quality, we conduct a focused evaluation on intrinsic retrieval metrics. Figure~\ref{tab:retrieval-results} details the comparative performance. Note that due to the lack of ground-truth ranking annotations in the Deep Research dataset, this analysis is restricted to benchmarks where gold labels are available.

We find MemSifter consistently achieves higher recall and ranking precision compared to embedding-based, hierarchical, and even reasoning-heavy retrieval baselines. By accurately filtering noise and prioritizing key evidence, our proxy ensures the working LLM receives the most relevant context. This high-quality retrieval acts as the foundational guarantee for the superior downstream performance reported in the previous section, confirming that precise memory recall is a prerequisite for effective reasoning.

\begin{figure}
    \centering
    \includegraphics[width=0.9\linewidth]{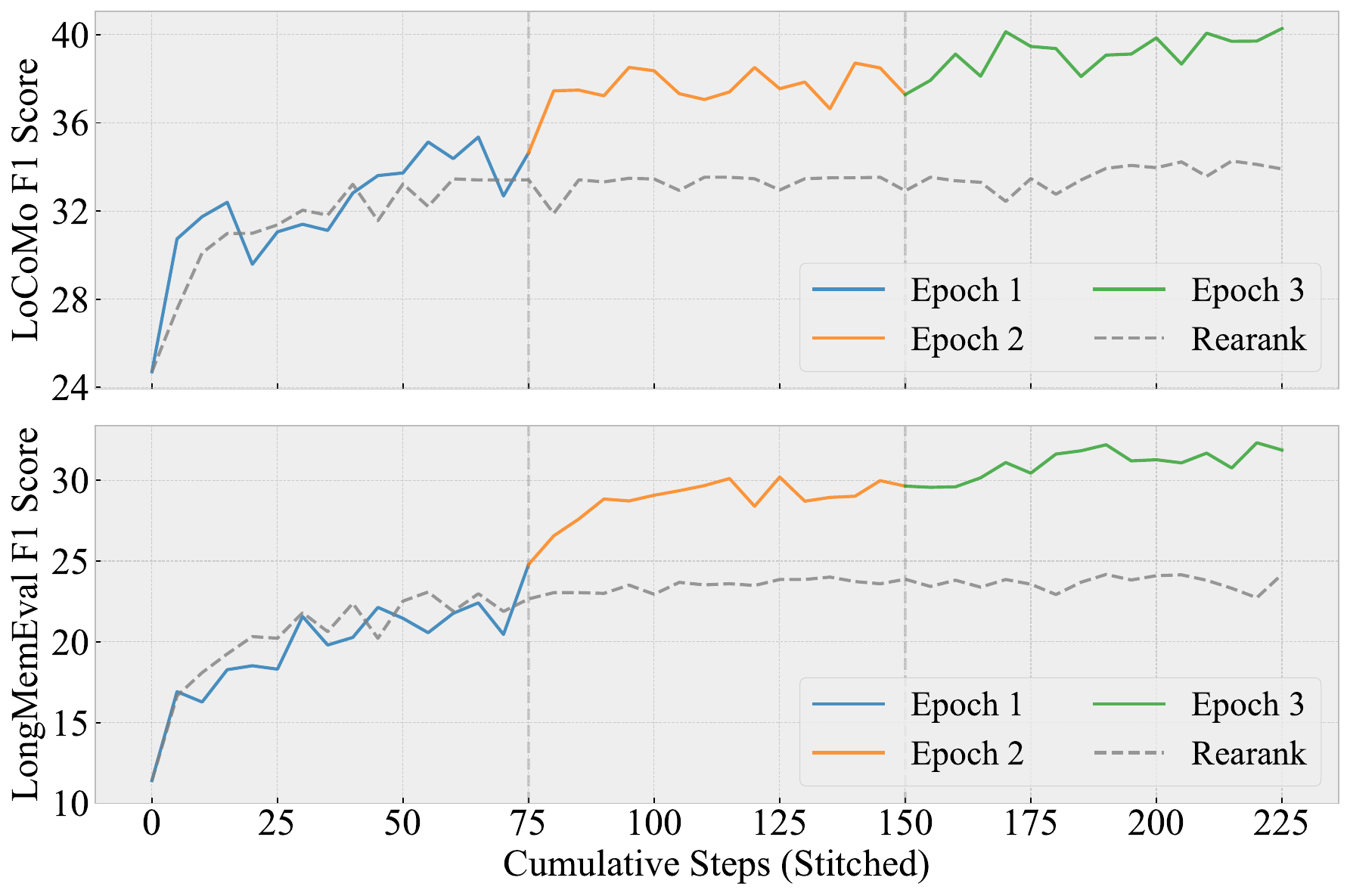}
    \caption{RL training curves of MemSifter compared with Rearank on two datasets~(Datasets Aligned).}
    \Description{Line plots comparing reinforcement learning training curves for MemSifter and Rearank on two aligned datasets.}
    \label{fig:rl-train}
\end{figure}

\subsubsection{\textbf{Training Dynamics and Curriculum Analysis}}

To assess our reward mechanism and iterative strategy, we plot the reinforcement learning trajectories in Figure~\ref{fig:rl-train}. The grey curve is the baseline (Rearank), and the colored segments show the three stages of MemSifter's iterative training. We conduct this analysis on LoCoMo and LongMemEval, which offer ground-truth rankings for baseline comparison. Two key observations validate our design:

\textbf{(1) Superior Alignment via Task-Outcome Reward:} 
MemSifter converges faster and achieves higher final performance than the baseline. While Rearank relies on static relevance labels, our \textbf{Task-Outcome-Oriented Reward} directly aligns the proxy with the working LLM’s end-to-end success. The steeper learning curve shows that rewarding the proxy for selecting memories that improve reasoning (via rank-sensitive gradients) is more effective than optimizing semantic similarity alone.

\textbf{(2) Breaking Plateaus with Curriculum Learning:} 
The baseline plateaus or degrades after 60–70 epochs~(grey trajectory), indicating it has exhausted the signal in
static data. In contrast, MemSifter continues to improve across three phases. By continually refreshing training data
to match the model’s evolving ability~(emphasizing samples near the difficulty anchor $\tau$) and using model merging,
we mitigate overfitting and achieves a higher upper bound.

\subsubsection{\textbf{Efficiency Analysis}}

We assess efficiency by normalizing input/output costs into \textbf{7B-equivalent token counts} based on model size
and by measuring single-H20 GPU latency~(Table~\ref{tab:efficiency-analysis}). Training cost is reported with the
same accounting fields in Appendix~\ref{sec:appendix-experimental-details}. The results show MemSifter’s key
trade-offs: (1) Relative to lightweight embedding baselines, it adds moderate latency from generation,
offset by large gains in retrieval precision. (2) Compared to “reason-then-rerank” methods, it achieves similar
computational cost with better task performance. (3) Most importantly, versus Long-Context LLMs over full histories,
MemSifter dramatically cuts overhead by delegating memory sifting to a smaller proxy, avoiding the heavy cost of
large models on long sequences~(128K$\rightarrow$2K).

\begin{table}[!tbp]
\small
\centering
\caption{Analysis of inference cost on WebDancer(128K).}
\label{tab:efficiency-analysis}
\setlength{\tabcolsep}{1.2mm}{
\begin{tabular}{lccrr}
\toprule
\textbf{Method}    & \textbf{\# Params} & \textbf{\# In-Tok} & \textbf{\# Out-Tok} & \textbf{Latency} \\
\midrule
BGE              & 0.2B   & 128K   & 61.08    & 1015.05ms   \\ 
ReaRank          & 7B     & 128K   & 1531.41  & 7657.05ms   \\ 
ReasonRank       & 7B     & 128K   & 1664.48  & 8322.40ms    \\ 
MemSifter        & 4B     & 128K   & 1557.01  & 3982.53ms   \\
DeepSeek-V3.2    & 632B   & 128K   & 712.48   & 49873.60ms   \\
DeepSeek-V3.2    & 632B   & 2K     & 705.61   & 759.89ms     \\
\bottomrule
\end{tabular}
}

\end{table}

\section{Conclusion and Future Work}

This paper tackles the trade-off between retrieval precision and computational efficiency in LLM memory systems. We introduce MemSifter, a framework that separates memory management from core reasoning by offloading historical interaction “sifting” to a lightweight proxy. Using a task-outcome-based reward—combining marginal utility and rank sensitivity, the proxy learns retrieval strategies from final task success alone, avoiding costly annotations. Experiments show MemSifter surpasses state-of-the-art baselines in task accuracy and inference efficiency. Our results present a scalable paradigm for long-horizon LLMs, where small specialized models curate context for large generalist reasoners. Future work will extend outcome-driven optimization to LLM memory consolidation and multi-modal histories.


\bibliographystyle{ACM-Reference-Format}

\begin{thebibliography}{60}


\ifx \showCODEN    \undefined \def \showCODEN     #1{\unskip}     \fi
\ifx \showISBNx    \undefined \def \showISBNx     #1{\unskip}     \fi
\ifx \showISBNxiii \undefined \def \showISBNxiii  #1{\unskip}     \fi
\ifx \showISSN     \undefined \def \showISSN      #1{\unskip}     \fi
\ifx \showLCCN     \undefined \def \showLCCN      #1{\unskip}     \fi
\ifx \shownote     \undefined \def \shownote      #1{#1}          \fi
\ifx \showarticletitle \undefined \def \showarticletitle #1{#1}   \fi
\ifx \showURL      \undefined \def \showURL       {\relax}        \fi
\providecommand\bibfield[2]{#2}
\providecommand\bibinfo[2]{#2}
\providecommand\natexlab[1]{#1}
\providecommand\showeprint[2][]{arXiv:#2}

\bibitem[Busolin et~al\mbox{.}(2025)]%
        {EfficientReranking}
\bibfield{author}{\bibinfo{person}{Francesco Busolin}, \bibinfo{person}{Claudio Lucchese}, \bibinfo{person}{Franco~Maria Nardini}, \bibinfo{person}{Salvatore Orlando}, \bibinfo{person}{Raffaele Perego}, \bibinfo{person}{Salvatore Trani}, {and} \bibinfo{person}{Alberto Veneri}.} \bibinfo{year}{2025}\natexlab{}.
\newblock \showarticletitle{Efficient Re-ranking with Cross-encoders via Early Exit}. In \bibinfo{booktitle}{\emph{Proceedings of the 48th International {ACM} {SIGIR} Conference on Research and Development in Information Retrieval, {SIGIR} 2025, Padua, Italy, July 13-18, 2025}}, \bibfield{editor}{\bibinfo{person}{Nicola Ferro}, \bibinfo{person}{Maria Maistro}, \bibinfo{person}{Gabriella Pasi}, \bibinfo{person}{Omar Alonso}, \bibinfo{person}{Andrew Trotman}, {and} \bibinfo{person}{Suzan Verberne}} (Eds.). \bibinfo{publisher}{{ACM}}, \bibinfo{pages}{2534--2544}.
\newblock
\href{https://doi.org/10.1145/3726302.3729962}{doi:\nolinkurl{10.1145/3726302.3729962}}


\bibitem[Chen et~al\mbox{.}(2024)]%
        {BGEM3}
\bibfield{author}{\bibinfo{person}{Jianlyu Chen}, \bibinfo{person}{Shitao Xiao}, \bibinfo{person}{Peitian Zhang}, \bibinfo{person}{Kun Luo}, \bibinfo{person}{Defu Lian}, {and} \bibinfo{person}{Zheng Liu}.} \bibinfo{year}{2024}\natexlab{}.
\newblock \showarticletitle{M3-Embedding: Multi-Linguality, Multi-Functionality, Multi-Granularity Text Embeddings Through Self-Knowledge Distillation}. In \bibinfo{booktitle}{\emph{Findings of the Association for Computational Linguistics, {ACL} 2024, Bangkok, Thailand and virtual meeting, August 11-16, 2024}}, \bibfield{editor}{\bibinfo{person}{Lun{-}Wei Ku}, \bibinfo{person}{Andre Martins}, {and} \bibinfo{person}{Vivek Srikumar}} (Eds.). \bibinfo{publisher}{Association for Computational Linguistics}, \bibinfo{pages}{2318--2335}.
\newblock
\href{https://doi.org/10.18653/V1/2024.FINDINGS-ACL.137}{doi:\nolinkurl{10.18653/V1/2024.FINDINGS-ACL.137}}


\bibitem[Chevalier et~al\mbox{.}(2023)]%
        {compress-context}
\bibfield{author}{\bibinfo{person}{Alexis Chevalier}, \bibinfo{person}{Alexander Wettig}, \bibinfo{person}{Anirudh Ajith}, {and} \bibinfo{person}{Danqi Chen}.} \bibinfo{year}{2023}\natexlab{}.
\newblock \showarticletitle{Adapting Language Models to Compress Contexts}. In \bibinfo{booktitle}{\emph{Proceedings of the 2023 Conference on Empirical Methods in Natural Language Processing, {EMNLP} 2023, Singapore, December 6-10, 2023}}, \bibfield{editor}{\bibinfo{person}{Houda Bouamor}, \bibinfo{person}{Juan Pino}, {and} \bibinfo{person}{Kalika Bali}} (Eds.). \bibinfo{publisher}{Association for Computational Linguistics}, \bibinfo{pages}{3829--3846}.
\newblock
\href{https://doi.org/10.18653/V1/2023.EMNLP-MAIN.232}{doi:\nolinkurl{10.18653/V1/2023.EMNLP-MAIN.232}}


\bibitem[Chhikara et~al\mbox{.}(2025)]%
        {mem0}
\bibfield{author}{\bibinfo{person}{Prateek Chhikara}, \bibinfo{person}{Dev Khant}, \bibinfo{person}{Saket Aryan}, \bibinfo{person}{Taranjeet Singh}, {and} \bibinfo{person}{Deshraj Yadav}.} \bibinfo{year}{2025}\natexlab{}.
\newblock \showarticletitle{Mem0: Building Production-Ready {AI} Agents with Scalable Long-Term Memory}.
\newblock \bibinfo{journal}{\emph{CoRR}}  \bibinfo{volume}{abs/2504.19413} (\bibinfo{year}{2025}).
\newblock
\showeprint[arXiv]{2504.19413}
\href{https://doi.org/10.48550/ARXIV.2504.19413}{doi:\nolinkurl{10.48550/ARXIV.2504.19413}}


\bibitem[Cui et~al\mbox{.}(2025)]%
        {TTE}
\bibfield{author}{\bibinfo{person}{Xuanming Cui}, \bibinfo{person}{Jianpeng Cheng}, \bibinfo{person}{Hong{-}you Chen}, \bibinfo{person}{Satya~Narayan Shukla}, \bibinfo{person}{Abhijeet Awasthi}, \bibinfo{person}{Xichen Pan}, \bibinfo{person}{Chaitanya Ahuja}, \bibinfo{person}{Shlok~Kumar Mishra}, \bibinfo{person}{Yonghuan Yang}, \bibinfo{person}{Jun Xiao}, \bibinfo{person}{Qi Guo}, \bibinfo{person}{Ser{-}Nam Lim}, \bibinfo{person}{Aashu Singh}, {and} \bibinfo{person}{Xiangjun Fan}.} \bibinfo{year}{2025}\natexlab{}.
\newblock \showarticletitle{Think Then Embed: Generative Context Improves Multimodal Embedding}.
\newblock \bibinfo{journal}{\emph{CoRR}}  \bibinfo{volume}{abs/2510.05014} (\bibinfo{year}{2025}).
\newblock
\showeprint[arXiv]{2510.05014}
\href{https://doi.org/10.48550/ARXIV.2510.05014}{doi:\nolinkurl{10.48550/ARXIV.2510.05014}}


\bibitem[DeepSeek-AI et~al\mbox{.}(2025)]%
        {DeepSeekV3.2}
\bibfield{author}{\bibinfo{person}{DeepSeek-AI}, \bibinfo{person}{Aixin Liu}, \bibinfo{person}{Aoxue Mei}, \bibinfo{person}{Bangcai Lin}, \bibinfo{person}{Bing Xue}, \bibinfo{person}{Bing-Li Wang}, \bibinfo{person}{Bingzheng Xu}, \bibinfo{person}{Bochao Wu}, \bibinfo{person}{Bowei Zhang}, \bibinfo{person}{Chaofan Lin}, \bibinfo{person}{Chen Dong}, \bibinfo{person}{Chengda Lu}, \bibinfo{person}{Chenggang Zhao}, \bibinfo{person}{Chengqi Deng}, \bibinfo{person}{Chenhao Xu}, \bibinfo{person}{Chong Ruan}, \bibinfo{person}{Damai Dai}, \bibinfo{person}{Daya Guo}, \bibinfo{person}{Dejian Yang}, \bibinfo{person}{Deli Chen}, \bibinfo{person}{Erhang Li}, \bibinfo{person}{Fangqi Zhou}, \bibinfo{person}{Fangyun Lin}, \bibinfo{person}{Fucong Dai}, \bibinfo{person}{Guangbo Hao}, \bibinfo{person}{Guanting Chen}, \bibinfo{person}{Guowei Li}, \bibinfo{person}{H. Zhang}, \bibinfo{person}{Hanwei Xu}, \bibinfo{person}{Hao Li}, \bibinfo{person}{Haofen Liang}, \bibinfo{person}{Haoran Wei}, \bibinfo{person}{Haowei Zhang},
  \bibinfo{person}{Hao sheng Luo}, \bibinfo{person}{Haozhe Ji}, \bibinfo{person}{Honghui Ding}, \bibinfo{person}{Hongxuan Tang}, \bibinfo{person}{Huan Cao}, \bibinfo{person}{Huazuo Gao}, \bibinfo{person}{Huixian Qu}, \bibinfo{person}{Hui Zeng}, \bibinfo{person}{Jialiang Huang}, \bibinfo{person}{Jiashi Li}, \bibinfo{person}{Jiaxin Xu}, \bibinfo{person}{Jiewen Hu}, \bibinfo{person}{JingChang Chen}, \bibinfo{person}{Jingting Xiang}, \bibinfo{person}{Jingyang Yuan}, \bibinfo{person}{Jing Cheng}, \bibinfo{person}{Jinhua Zhu}, \bibinfo{person}{Jun Ran}, \bibinfo{person}{Junguang Jiang}, \bibinfo{person}{Junjie Qiu}, \bibinfo{person}{Junlong Li}, \bibinfo{person}{Jun-Mei Song}, \bibinfo{person}{Kai Dong}, \bibinfo{person}{Kaige Gao}, \bibinfo{person}{Kang Guan}, \bibinfo{person}{Kexin Huang}, \bibinfo{person}{Kexing Zhou}, \bibinfo{person}{Kezhao Huang}, \bibinfo{person}{Kuai Yu}, \bibinfo{person}{Lean Wang}, \bibinfo{person}{Lecong Zhang}, \bibinfo{person}{Lei Wang}, \bibinfo{person}{Liang Zhao},
  \bibinfo{person}{Liangsheng Yin}, \bibinfo{person}{Lihua Guo}, \bibinfo{person}{Ling-Li Luo}, \bibinfo{person}{Linwang Ma}, \bibinfo{person}{Litong Wang}, \bibinfo{person}{Liyue Zhang}, \bibinfo{person}{M.~S. Di}, \bibinfo{person}{M.~Y. Xu}, \bibinfo{person}{Mingchuan Zhang}, \bibinfo{person}{Minghua Zhang}, \bibinfo{person}{Min Tang}, \bibinfo{person}{Mingxu Zhou}, \bibinfo{person}{P. Huang}, \bibinfo{person}{Peixin Cong}, \bibinfo{person}{Peiyi Wang}, \bibinfo{person}{Qiancheng Wang}, \bibinfo{person}{Qihao Zhu}, \bibinfo{person}{Qingyang Li}, \bibinfo{person}{Qinyu Chen}, \bibinfo{person}{Qiushi Du}, \bibinfo{person}{Ruiling Xu}, \bibinfo{person}{Ruiqi Ge}, \bibinfo{person}{Ruisong Zhang}, \bibinfo{person}{Ruizhe Pan}, \bibinfo{person}{Runji Wang}, \bibinfo{person}{Runqiu Yin}, \bibinfo{person}{Runxin Xu}, \bibinfo{person}{Ruomeng Shen}, \bibinfo{person}{Ruoyu Zhang}, \bibinfo{person}{S.~H. Liu}, \bibinfo{person}{Shanghao Lu}, \bibinfo{person}{Shangyan Zhou}, \bibinfo{person}{Shanhuang Chen},
  \bibinfo{person}{Shaofei Cai}, \bibinfo{person}{Shaoyuan Chen}, \bibinfo{person}{Shengding Hu}, \bibinfo{person}{Shengyu Liu}, \bibinfo{person}{Shiqiang Hu}, \bibinfo{person}{Shirong Ma}, \bibinfo{person}{Shiyu Wang}, \bibinfo{person}{Shuiping Yu}, \bibinfo{person}{Shunfeng Zhou}, \bibinfo{person}{Shuting Pan}, \bibinfo{person}{Songyang Zhou}, \bibinfo{person}{Tao Ni}, \bibinfo{person}{Tao Yun}, \bibinfo{person}{Tian Pei}, \bibinfo{person}{Tian Ye}, \bibinfo{person}{Tianyuan Yue}, \bibinfo{person}{Wangding Zeng}, \bibinfo{person}{Wen Liu}, \bibinfo{person}{Wenfeng Liang}, \bibinfo{person}{Wenjie Pang}, \bibinfo{person}{Wenjing Luo}, \bibinfo{person}{Wenjun Gao}, \bibinfo{person}{Wentao Zhang}, \bibinfo{person}{Xi Gao}, \bibinfo{person}{Xiangwen Wang}, \bibinfo{person}{Xiaoling Bi}, \bibinfo{person}{Xiaodong Liu}, \bibinfo{person}{Xiaohan Wang}, \bibinfo{person}{Xiaokang Chen}, \bibinfo{person}{Xiaokang Zhang}, \bibinfo{person}{Xiaotao Nie}, \bibinfo{person}{Xin Cheng}, \bibinfo{person}{Xin Liu},
  \bibinfo{person}{Xin Xie}, \bibinfo{person}{Xingchao Liu}, \bibinfo{person}{Xingkai Yu}, \bibinfo{person}{Xingyou Li}, \bibinfo{person}{Xinyu Yang}, \bibinfo{person}{Xinyuan Li}, \bibinfo{person}{Xu Chen}, \bibinfo{person}{Xuecheng Su}, \bibinfo{person}{Xuehai Pan}, \bibinfo{person}{Xuheng Lin}, \bibinfo{person}{Xuwei Fu}, \bibinfo{person}{Y.~Q. Wang}, \bibinfo{person}{Yang Zhang}, \bibinfo{person}{Yanhong Xu}, \bibinfo{person}{Yanru Ma}, \bibinfo{person}{Yao Li}, \bibinfo{person}{Yao Zhao}, \bibinfo{person}{Yaofeng Sun}, \bibinfo{person}{Yaohui Wang}, \bibinfo{person}{Yi Qian}, \bibinfo{person}{Yingpu Yu}, \bibinfo{person}{Yichao Zhang}, \bibinfo{person}{Yifan Ding}, \bibinfo{person}{Yifan Shi}, \bibinfo{person}{Yi Xiong}, \bibinfo{person}{Ying He}, \bibinfo{person}{Ying Zhou}, \bibinfo{person}{Yinmin Zhong}, \bibinfo{person}{Yishi Piao}, \bibinfo{person}{Yisong Wang}, \bibinfo{person}{Yixiao Chen}, \bibinfo{person}{Yixuan Tan}, \bibinfo{person}{Yixuan Wei}, \bibinfo{person}{Yiyang Ma},
  \bibinfo{person}{Yiyuan Liu}, \bibinfo{person}{Yonglun Yang}, \bibinfo{person}{Yongqiang Guo}, \bibinfo{person}{Yongtong Wu}, \bibinfo{person}{Yu Wu}, \bibinfo{person}{Yuan Cheng}, \bibinfo{person}{Yuan Ou}, \bibinfo{person}{Yuanfan Xu}, \bibinfo{person}{Yuduan Wang}, \bibinfo{person}{Yue Gong}, \bibinfo{person}{Yuhan Wu}, \bibinfo{person}{Yuheng Zou}, \bibinfo{person}{Yukun Li}, \bibinfo{person}{Yunfan Xiong}, \bibinfo{person}{Yu-Wei Luo}, \bibinfo{person}{Yu mei You}, \bibinfo{person}{Yuxuan Liu}, \bibinfo{person}{Yuyang Zhou}, \bibinfo{person}{Z.~F. Wu}, \bibinfo{person}{Zehui Ren}, \bibinfo{person}{Zehua Zhao}, \bibinfo{person}{Zehui Ren}, \bibinfo{person}{Zhangli Sha}, \bibinfo{person}{Zhe Fu}, \bibinfo{person}{Zhean Xu}, \bibinfo{person}{Zhenda Xie}, \bibinfo{person}{Zhen guo Zhang}, \bibinfo{person}{Zhewen Hao}, \bibinfo{person}{Zhibin Gou}, \bibinfo{person}{Zhicheng Ma}, \bibinfo{person}{Zhigang Yan}, \bibinfo{person}{Zhihong Shao}, \bibinfo{person}{Zhixian Huang}, \bibinfo{person}{Zhiyu Wu},
  \bibinfo{person}{Zhuoshu Li}, \bibinfo{person}{Zhuping Zhang}, \bibinfo{person}{Zian Xu}, \bibinfo{person}{Zihao Wang}, \bibinfo{person}{Zihui Gu}, \bibinfo{person}{Zijia Zhu}, \bibinfo{person}{Zi-Rui Li}, \bibinfo{person}{Zipeng Zhang}, \bibinfo{person}{Ziwei Xie}, \bibinfo{person}{Ziyi Gao}, \bibinfo{person}{Zizheng Pan}, \bibinfo{person}{Zongqing Yao}, \bibinfo{person}{Bei Feng}, \bibinfo{person}{Hui Li}, \bibinfo{person}{J.~L. Cai}, \bibinfo{person}{Jiaqi Ni}, \bibinfo{person}{Lei Xu}, \bibinfo{person}{Meng Li}, \bibinfo{person}{Ning Tian}, \bibinfo{person}{R.~J. Chen}, \bibinfo{person}{Ruiqi Jin}, \bibinfo{person}{S.~S. Li}, \bibinfo{person}{Shuang Zhou}, \bibinfo{person}{Tianyu Sun}, \bibinfo{person}{X.~Q. Li}, \bibinfo{person}{Xiangyu Jin}, \bibinfo{person}{Xiaojin Shen}, \bibinfo{person}{Xiaosha Chen}, \bibinfo{person}{Xinnan Song}, \bibinfo{person}{Xinyi Zhou}, \bibinfo{person}{Y.~X. Zhu}, \bibinfo{person}{Yanping Huang}, \bibinfo{person}{Yao Li}, \bibinfo{person}{Yi Zheng}, \bibinfo{person}{Yuchen
  Zhu}, \bibinfo{person}{Yunxiang Ma}, \bibinfo{person}{Zhen Huang}, \bibinfo{person}{Zhipeng Xu}, \bibinfo{person}{Zhongyu Zhang}, \bibinfo{person}{Dong-Li Ji}, \bibinfo{person}{Jian Liang}, \bibinfo{person}{Jianzhong Guo}, \bibinfo{person}{Jin Chen}, \bibinfo{person}{Leyi Xia}, \bibinfo{person}{Miaojun Wang}, \bibinfo{person}{Mingming Li}, \bibinfo{person}{Peng Zhang}, \bibinfo{person}{Ruyi Chen}, \bibinfo{person}{Shangmian Sun}, \bibinfo{person}{Shao-Kang Wu}, \bibinfo{person}{Shengfeng Ye}, \bibinfo{person}{T.Wang}, \bibinfo{person}{W.~L. Xiao}, \bibinfo{person}{Wei An}, \bibinfo{person}{Xianzu Wang}, \bibinfo{person}{Xiaowen Sun}, \bibinfo{person}{Xiaoxiang Wang}, \bibinfo{person}{Ying Tang}, \bibinfo{person}{Yukun Zha}, \bibinfo{person}{Ze-Na Zhang}, \bibinfo{person}{Zhenghua Ju}, \bibinfo{person}{Zhen Zhang}, {and} \bibinfo{person}{Zihua Qu}.} \bibinfo{year}{2025}\natexlab{}.
\newblock \showarticletitle{DeepSeek-V3.2: Pushing the Frontier of Open Large Language Models}.
\newblock
\urldef\tempurl%
\url{https://api.semanticscholar.org/CorpusID:283448719}
\showURL{%
\tempurl}


\bibitem[Du et~al\mbox{.}(2024)]%
        {PerLTQA}
\bibfield{author}{\bibinfo{person}{Yiming Du}, \bibinfo{person}{Hongru Wang}, \bibinfo{person}{Zhengyi Zhao}, \bibinfo{person}{Bin Liang}, \bibinfo{person}{Baojun Wang}, \bibinfo{person}{Wanjun Zhong}, \bibinfo{person}{Zezhong Wang}, {and} \bibinfo{person}{Kam{-}Fai Wong}.} \bibinfo{year}{2024}\natexlab{}.
\newblock \showarticletitle{PerLTQA: {A} Personal Long-Term Memory Dataset for Memory Classification, Retrieval, and Synthesis in Question Answering}.
\newblock \bibinfo{journal}{\emph{CoRR}}  \bibinfo{volume}{abs/2402.16288} (\bibinfo{year}{2024}).
\newblock
\showeprint[arXiv]{2402.16288}
\href{https://doi.org/10.48550/ARXIV.2402.16288}{doi:\nolinkurl{10.48550/ARXIV.2402.16288}}


\bibitem[Edge et~al\mbox{.}(2024)]%
        {graphrag}
\bibfield{author}{\bibinfo{person}{Darren Edge}, \bibinfo{person}{Ha Trinh}, \bibinfo{person}{Newman Cheng}, \bibinfo{person}{Joshua Bradley}, \bibinfo{person}{Alex Chao}, \bibinfo{person}{Apurva Mody}, \bibinfo{person}{Steven Truitt}, {and} \bibinfo{person}{Jonathan Larson}.} \bibinfo{year}{2024}\natexlab{}.
\newblock \showarticletitle{From Local to Global: {A} Graph {RAG} Approach to Query-Focused Summarization}.
\newblock \bibinfo{journal}{\emph{CoRR}}  \bibinfo{volume}{abs/2404.16130} (\bibinfo{year}{2024}).
\newblock
\showeprint[arXiv]{2404.16130}
\href{https://doi.org/10.48550/ARXIV.2404.16130}{doi:\nolinkurl{10.48550/ARXIV.2404.16130}}


\bibitem[Gutierrez et~al\mbox{.}(2024)]%
        {HippoRAG}
\bibfield{author}{\bibinfo{person}{Bernal~Jimenez Gutierrez}, \bibinfo{person}{Yiheng Shu}, \bibinfo{person}{Yu Gu}, \bibinfo{person}{Michihiro Yasunaga}, {and} \bibinfo{person}{Yu Su}.} \bibinfo{year}{2024}\natexlab{}.
\newblock \showarticletitle{HippoRAG: Neurobiologically Inspired Long-Term Memory for Large Language Models}. In \bibinfo{booktitle}{\emph{Advances in Neural Information Processing Systems 38: Annual Conference on Neural Information Processing Systems 2024, NeurIPS 2024, Vancouver, BC, Canada, December 10 - 15, 2024}}, \bibfield{editor}{\bibinfo{person}{Amir Globersons}, \bibinfo{person}{Lester Mackey}, \bibinfo{person}{Danielle Belgrave}, \bibinfo{person}{Angela Fan}, \bibinfo{person}{Ulrich Paquet}, \bibinfo{person}{Jakub~M. Tomczak}, {and} \bibinfo{person}{Cheng Zhang}} (Eds.).
\newblock
\urldef\tempurl%
\url{http://papers.nips.cc/paper\_files/paper/2024/hash/6ddc001d07ca4f319af96a3024f6dbd1-Abstract-Conference.html}
\showURL{%
\tempurl}


\bibitem[Hu et~al\mbox{.}(2024)]%
        {RankPrompt}
\bibfield{author}{\bibinfo{person}{Chi Hu}, \bibinfo{person}{Yuan Ge}, \bibinfo{person}{Xiangnan Ma}, \bibinfo{person}{Hang Cao}, \bibinfo{person}{Qiang Li}, \bibinfo{person}{Yonghua Yang}, \bibinfo{person}{Tong Xiao}, {and} \bibinfo{person}{Jingbo Zhu}.} \bibinfo{year}{2024}\natexlab{}.
\newblock \showarticletitle{RankPrompt: Step-by-Step Comparisons Make Language Models Better Reasoners}. In \bibinfo{booktitle}{\emph{Proceedings of the 2024 Joint International Conference on Computational Linguistics, Language Resources and Evaluation, {LREC/COLING} 2024, 20-25 May, 2024, Torino, Italy}}, \bibfield{editor}{\bibinfo{person}{Nicoletta Calzolari}, \bibinfo{person}{Min{-}Yen Kan}, \bibinfo{person}{V{\'{e}}ronique Hoste}, \bibinfo{person}{Alessandro Lenci}, \bibinfo{person}{Sakriani Sakti}, {and} \bibinfo{person}{Nianwen Xue}} (Eds.). \bibinfo{publisher}{{ELRA} and {ICCL}}, \bibinfo{pages}{13524--13536}.
\newblock
\urldef\tempurl%
\url{https://aclanthology.org/2024.lrec-main.1183}
\showURL{%
\tempurl}


\bibitem[Hu et~al\mbox{.}(2026)]%
        {memory-matters}
\bibfield{author}{\bibinfo{person}{Yuyang Hu}, \bibinfo{person}{Jiongnan Liu}, \bibinfo{person}{Jiejun Tan}, \bibinfo{person}{Yutao Zhu}, {and} \bibinfo{person}{Zhicheng Dou}.} \bibinfo{year}{2026}\natexlab{}.
\newblock \showarticletitle{Memory Matters More: Event-Centric Memory as a Logic Map for Agent Searching and Reasoning}.
\newblock  (\bibinfo{year}{2026}).
\newblock
\showeprint[arxiv]{2601.04726}~[cs.AI]
\urldef\tempurl%
\url{https://arxiv.org/abs/2601.04726}
\showURL{%
\tempurl}


\bibitem[Hu et~al\mbox{.}(2025)]%
        {mem-survey}
\bibfield{author}{\bibinfo{person}{Yuyang Hu}, \bibinfo{person}{Shichun Liu}, \bibinfo{person}{Yanwei Yue}, \bibinfo{person}{Guibin Zhang}, \bibinfo{person}{Boyang Liu}, \bibinfo{person}{Fangyi Zhu}, \bibinfo{person}{Jiahang Lin}, \bibinfo{person}{Honglin Guo}, \bibinfo{person}{Shihan Dou}, \bibinfo{person}{Zhiheng Xi}, \bibinfo{person}{Senjie Jin}, \bibinfo{person}{Jiejun Tan}, \bibinfo{person}{Yanbin Yin}, \bibinfo{person}{Jiongnan Liu}, \bibinfo{person}{Zeyu Zhang}, \bibinfo{person}{Zhongxiang Sun}, \bibinfo{person}{Yutao Zhu}, \bibinfo{person}{Hao Sun}, \bibinfo{person}{Boci Peng}, \bibinfo{person}{Zhenrong Cheng}, \bibinfo{person}{Xuanbo Fan}, \bibinfo{person}{Jiaxin Guo}, \bibinfo{person}{Xinlei Yu}, \bibinfo{person}{Zhenhong Zhou}, \bibinfo{person}{Zewen Hu}, \bibinfo{person}{Jiahao Huo}, \bibinfo{person}{Junhao Wang}, \bibinfo{person}{Yuwei Niu}, \bibinfo{person}{Yu Wang}, \bibinfo{person}{Zhenfei Yin}, \bibinfo{person}{Xiaobin Hu}, \bibinfo{person}{Yue Liao}, \bibinfo{person}{Qiankun Li},
  \bibinfo{person}{Kun Wang}, \bibinfo{person}{Wangchunshu Zhou}, \bibinfo{person}{Yixin Liu}, \bibinfo{person}{Dawei Cheng}, \bibinfo{person}{Qi Zhang}, \bibinfo{person}{Tao Gui}, \bibinfo{person}{Shirui Pan}, \bibinfo{person}{Yan Zhang}, \bibinfo{person}{Philip Torr}, \bibinfo{person}{Zhicheng Dou}, \bibinfo{person}{Ji-Rong Wen}, \bibinfo{person}{Xuanjing Huang}, \bibinfo{person}{Yu-Gang Jiang}, {and} \bibinfo{person}{Shuicheng Yan}.} \bibinfo{year}{2025}\natexlab{}.
\newblock \bibinfo{title}{Memory in the Age of AI Agents}.
\newblock
\showeprint[arxiv]{2512.13564}~[cs.CL]
\urldef\tempurl%
\url{https://arxiv.org/abs/2512.13564}
\showURL{%
\tempurl}


\bibitem[Izacard et~al\mbox{.}(2023)]%
        {atlas}
\bibfield{author}{\bibinfo{person}{Gautier Izacard}, \bibinfo{person}{Patrick Lewis}, \bibinfo{person}{Maria Lomeli}, \bibinfo{person}{Lucas Hosseini}, \bibinfo{person}{Fabio Petroni}, \bibinfo{person}{Timo Schick}, \bibinfo{person}{Jane Dwivedi{-}Yu}, \bibinfo{person}{Armand Joulin}, \bibinfo{person}{Sebastian Riedel}, {and} \bibinfo{person}{Edouard Grave}.} \bibinfo{year}{2023}\natexlab{}.
\newblock \showarticletitle{Atlas: Few-shot Learning with Retrieval Augmented Language Models}.
\newblock \bibinfo{journal}{\emph{J. Mach. Learn. Res.}}  \bibinfo{volume}{24} (\bibinfo{year}{2023}), \bibinfo{pages}{251:1--251:43}.
\newblock
\urldef\tempurl%
\url{https://jmlr.org/papers/v24/23-0037.html}
\showURL{%
\tempurl}


\bibitem[J{\"{a}}rvelin and Kek{\"{a}}l{\"{a}}inen(2002)]%
        {dcg}
\bibfield{author}{\bibinfo{person}{Kalervo J{\"{a}}rvelin} {and} \bibinfo{person}{Jaana Kek{\"{a}}l{\"{a}}inen}.} \bibinfo{year}{2002}\natexlab{}.
\newblock \showarticletitle{Cumulated gain-based evaluation of {IR} techniques}.
\newblock \bibinfo{journal}{\emph{{ACM} Trans. Inf. Syst.}} \bibinfo{volume}{20}, \bibinfo{number}{4} (\bibinfo{year}{2002}), \bibinfo{pages}{422--446}.
\newblock
\href{https://doi.org/10.1145/582415.582418}{doi:\nolinkurl{10.1145/582415.582418}}


\bibitem[Ji et~al\mbox{.}(2024)]%
        {memmap}
\bibfield{author}{\bibinfo{person}{Shuo Ji}, \bibinfo{person}{Mingzhe Liu}, \bibinfo{person}{Leilei Sun}, \bibinfo{person}{Chuanren Liu}, {and} \bibinfo{person}{Tongyu Zhu}.} \bibinfo{year}{2024}\natexlab{}.
\newblock \showarticletitle{MemMap: An Adaptive and Latent Memory Structure for Dynamic Graph Learning}. In \bibinfo{booktitle}{\emph{Proceedings of the 30th {ACM} {SIGKDD} Conference on Knowledge Discovery and Data Mining, {KDD} 2024, Barcelona, Spain, August 25-29, 2024}}, \bibfield{editor}{\bibinfo{person}{Ricardo Baeza{-}Yates} {and} \bibinfo{person}{Francesco Bonchi}} (Eds.). \bibinfo{publisher}{{ACM}}, \bibinfo{pages}{1257--1268}.
\newblock
\href{https://doi.org/10.1145/3637528.3672060}{doi:\nolinkurl{10.1145/3637528.3672060}}


\bibitem[Jiang et~al\mbox{.}(2025a)]%
        {personamem}
\bibfield{author}{\bibinfo{person}{Bowen Jiang}, \bibinfo{person}{Zhuoqun Hao}, \bibinfo{person}{Young{-}Min Cho}, \bibinfo{person}{Bryan Li}, \bibinfo{person}{Yuan Yuan}, \bibinfo{person}{Sihao Chen}, \bibinfo{person}{Lyle~H. Ungar}, \bibinfo{person}{Camillo~J. Taylor}, {and} \bibinfo{person}{Dan Roth}.} \bibinfo{year}{2025}\natexlab{a}.
\newblock \showarticletitle{Know Me, Respond to Me: Benchmarking LLMs for Dynamic User Profiling and Personalized Responses at Scale}.
\newblock \bibinfo{journal}{\emph{CoRR}}  \bibinfo{volume}{abs/2504.14225} (\bibinfo{year}{2025}).
\newblock
\showeprint[arXiv]{2504.14225}
\href{https://doi.org/10.48550/ARXIV.2504.14225}{doi:\nolinkurl{10.48550/ARXIV.2504.14225}}


\bibitem[Jiang et~al\mbox{.}(2025b)]%
        {Jiang2025PersonaMemv2TP}
\bibfield{author}{\bibinfo{person}{Bowen Jiang}, \bibinfo{person}{Yuan Yuan}, \bibinfo{person}{Maohao Shen}, \bibinfo{person}{Zhuoqun Hao}, \bibinfo{person}{Zhangchen Xu}, \bibinfo{person}{Zichen Chen}, \bibinfo{person}{Ziyi Liu}, \bibinfo{person}{Anvesh~Rao Vijjini}, \bibinfo{person}{Jiashu He}, \bibinfo{person}{Hanchao Yu}, \bibinfo{person}{Radha Poovendran}, \bibinfo{person}{Greg Wornell}, \bibinfo{person}{Lyle Ungar}, \bibinfo{person}{Dan Roth}, \bibinfo{person}{Sihao Chen}, {and} \bibinfo{person}{Camillo~Jose Taylor}.} \bibinfo{year}{2025}\natexlab{b}.
\newblock \showarticletitle{PersonaMem-v2: Towards Personalized Intelligence via Learning Implicit User Personas and Agentic Memory}.
\newblock
\urldef\tempurl%
\url{https://api.semanticscholar.org/CorpusID:283693901}
\showURL{%
\tempurl}


\bibitem[Jimenez et~al\mbox{.}(2024)]%
        {swe-bench}
\bibfield{author}{\bibinfo{person}{Carlos~E. Jimenez}, \bibinfo{person}{John Yang}, \bibinfo{person}{Alexander Wettig}, \bibinfo{person}{Shunyu Yao}, \bibinfo{person}{Kexin Pei}, \bibinfo{person}{Ofir Press}, {and} \bibinfo{person}{Karthik~R. Narasimhan}.} \bibinfo{year}{2024}\natexlab{}.
\newblock \showarticletitle{SWE-bench: Can Language Models Resolve Real-world Github Issues?}. In \bibinfo{booktitle}{\emph{The Twelfth International Conference on Learning Representations, {ICLR} 2024, Vienna, Austria, May 7-11, 2024}}. \bibinfo{publisher}{OpenReview.net}.
\newblock
\urldef\tempurl%
\url{https://openreview.net/forum?id=VTF8yNQM66}
\showURL{%
\tempurl}


\bibitem[Liang et~al\mbox{.}(2022)]%
        {HolisticEvaluationLMs}
\bibfield{author}{\bibinfo{person}{Percy Liang}, \bibinfo{person}{Rishi Bommasani}, \bibinfo{person}{Tony Lee}, \bibinfo{person}{Dimitris Tsipras}, \bibinfo{person}{Dilara Soylu}, \bibinfo{person}{Michihiro Yasunaga}, \bibinfo{person}{Yian Zhang}, \bibinfo{person}{Deepak Narayanan}, \bibinfo{person}{Yuhuai Wu}, \bibinfo{person}{Ananya Kumar}, \bibinfo{person}{Benjamin Newman}, \bibinfo{person}{Binhang Yuan}, \bibinfo{person}{Bobby Yan}, \bibinfo{person}{Ce Zhang}, \bibinfo{person}{Christian Cosgrove}, \bibinfo{person}{Christopher~D. Manning}, \bibinfo{person}{Christopher R{\'{e}}}, \bibinfo{person}{Diana Acosta{-}Navas}, \bibinfo{person}{Drew~A. Hudson}, \bibinfo{person}{Eric Zelikman}, \bibinfo{person}{Esin Durmus}, \bibinfo{person}{Faisal Ladhak}, \bibinfo{person}{Frieda Rong}, \bibinfo{person}{Hongyu Ren}, \bibinfo{person}{Huaxiu Yao}, \bibinfo{person}{Jue Wang}, \bibinfo{person}{Keshav Santhanam}, \bibinfo{person}{Laurel~J. Orr}, \bibinfo{person}{Lucia Zheng}, \bibinfo{person}{Mert
  Y{\"{u}}ksekg{\"{o}}n{\"{u}}l}, \bibinfo{person}{Mirac Suzgun}, \bibinfo{person}{Nathan Kim}, \bibinfo{person}{Neel Guha}, \bibinfo{person}{Niladri~S. Chatterji}, \bibinfo{person}{Omar Khattab}, \bibinfo{person}{Peter Henderson}, \bibinfo{person}{Qian Huang}, \bibinfo{person}{Ryan Chi}, \bibinfo{person}{Sang~Michael Xie}, \bibinfo{person}{Shibani Santurkar}, \bibinfo{person}{Surya Ganguli}, \bibinfo{person}{Tatsunori Hashimoto}, \bibinfo{person}{Thomas Icard}, \bibinfo{person}{Tianyi Zhang}, \bibinfo{person}{Vishrav Chaudhary}, \bibinfo{person}{William Wang}, \bibinfo{person}{Xuechen Li}, \bibinfo{person}{Yifan Mai}, \bibinfo{person}{Yuhui Zhang}, {and} \bibinfo{person}{Yuta Koreeda}.} \bibinfo{year}{2022}\natexlab{}.
\newblock \showarticletitle{Holistic Evaluation of Language Models}.
\newblock \bibinfo{journal}{\emph{CoRR}}  \bibinfo{volume}{abs/2211.09110} (\bibinfo{year}{2022}).
\newblock
\showeprint[arXiv]{2211.09110}
\href{https://doi.org/10.48550/ARXIV.2211.09110}{doi:\nolinkurl{10.48550/ARXIV.2211.09110}}


\bibitem[Liu et~al\mbox{.}(2025b)]%
        {Liu2025ReasoningGE}
\bibfield{author}{\bibinfo{person}{Chunxu Liu}, \bibinfo{person}{Jiyuan Yang}, \bibinfo{person}{Ruopeng Gao}, \bibinfo{person}{Yuhan Zhu}, \bibinfo{person}{Feng Zhu}, \bibinfo{person}{Rui Zhao}, {and} \bibinfo{person}{Limin Wang}.} \bibinfo{year}{2025}\natexlab{b}.
\newblock \showarticletitle{Reasoning Guided Embeddings: Leveraging MLLM Reasoning for Improved Multimodal Retrieval}.
\newblock
\urldef\tempurl%
\url{https://api.semanticscholar.org/CorpusID:283109952}
\showURL{%
\tempurl}


\bibitem[Liu et~al\mbox{.}(2025a)]%
        {reasonrank}
\bibfield{author}{\bibinfo{person}{Wenhan Liu}, \bibinfo{person}{Xinyu Ma}, \bibinfo{person}{Weiwei Sun}, \bibinfo{person}{Yutao Zhu}, \bibinfo{person}{Yuchen Li}, \bibinfo{person}{Dawei Yin}, {and} \bibinfo{person}{Zhicheng Dou}.} \bibinfo{year}{2025}\natexlab{a}.
\newblock \showarticletitle{ReasonRank: Empowering Passage Ranking with Strong Reasoning Ability}.
\newblock \bibinfo{journal}{\emph{CoRR}}  \bibinfo{volume}{abs/2508.07050} (\bibinfo{year}{2025}).
\newblock
\showeprint[arXiv]{2508.07050}
\href{https://doi.org/10.48550/ARXIV.2508.07050}{doi:\nolinkurl{10.48550/ARXIV.2508.07050}}


\bibitem[Maharana et~al\mbox{.}(2024)]%
        {locomo}
\bibfield{author}{\bibinfo{person}{Adyasha Maharana}, \bibinfo{person}{Dong{-}Ho Lee}, \bibinfo{person}{Sergey Tulyakov}, \bibinfo{person}{Mohit Bansal}, \bibinfo{person}{Francesco Barbieri}, {and} \bibinfo{person}{Yuwei Fang}.} \bibinfo{year}{2024}\natexlab{}.
\newblock \showarticletitle{Evaluating Very Long-Term Conversational Memory of {LLM} Agents}. In \bibinfo{booktitle}{\emph{Proceedings of the 62nd Annual Meeting of the Association for Computational Linguistics (Volume 1: Long Papers), {ACL} 2024, Bangkok, Thailand, August 11-16, 2024}}, \bibfield{editor}{\bibinfo{person}{Lun{-}Wei Ku}, \bibinfo{person}{Andre Martins}, {and} \bibinfo{person}{Vivek Srikumar}} (Eds.). \bibinfo{publisher}{Association for Computational Linguistics}, \bibinfo{pages}{13851--13870}.
\newblock
\href{https://doi.org/10.18653/V1/2024.ACL-LONG.747}{doi:\nolinkurl{10.18653/V1/2024.ACL-LONG.747}}


\bibitem[Mialon et~al\mbox{.}(2024)]%
        {gaia}
\bibfield{author}{\bibinfo{person}{Gr{\'{e}}goire Mialon}, \bibinfo{person}{Cl{\'{e}}mentine Fourrier}, \bibinfo{person}{Thomas Wolf}, \bibinfo{person}{Yann LeCun}, {and} \bibinfo{person}{Thomas Scialom}.} \bibinfo{year}{2024}\natexlab{}.
\newblock \showarticletitle{{GAIA:} a benchmark for General {AI} Assistants}. In \bibinfo{booktitle}{\emph{The Twelfth International Conference on Learning Representations, {ICLR} 2024, Vienna, Austria, May 7-11, 2024}}. \bibinfo{publisher}{OpenReview.net}.
\newblock
\urldef\tempurl%
\url{https://openreview.net/forum?id=fibxvahvs3}
\showURL{%
\tempurl}


\bibitem[Modarressi et~al\mbox{.}(2025)]%
        {memLLM}
\bibfield{author}{\bibinfo{person}{Ali Modarressi}, \bibinfo{person}{Abdullatif K{\"{o}}ksal}, \bibinfo{person}{Ayyoob Imani}, \bibinfo{person}{Mohsen Fayyaz}, {and} \bibinfo{person}{Hinrich Sch{\"{u}}tze}.} \bibinfo{year}{2025}\natexlab{}.
\newblock \showarticletitle{MemLLM: Finetuning LLMs to Use Explicit Read-Write Memory}.
\newblock \bibinfo{journal}{\emph{Trans. Mach. Learn. Res.}}  \bibinfo{volume}{2025} (\bibinfo{year}{2025}).
\newblock
\urldef\tempurl%
\url{https://openreview.net/forum?id=dghM7sOudh}
\showURL{%
\tempurl}


\bibitem[Nan et~al\mbox{.}(2025)]%
        {Nemori}
\bibfield{author}{\bibinfo{person}{Jiayan Nan}, \bibinfo{person}{Wenquan Ma}, \bibinfo{person}{Wenlong Wu}, {and} \bibinfo{person}{Yize Chen}.} \bibinfo{year}{2025}\natexlab{}.
\newblock \showarticletitle{Nemori: Self-Organizing Agent Memory Inspired by Cognitive Science}.
\newblock \bibinfo{journal}{\emph{CoRR}}  \bibinfo{volume}{abs/2508.03341} (\bibinfo{year}{2025}).
\newblock
\showeprint[arXiv]{2508.03341}
\href{https://doi.org/10.48550/ARXIV.2508.03341}{doi:\nolinkurl{10.48550/ARXIV.2508.03341}}


\bibitem[Packer et~al\mbox{.}(2023)]%
        {MemGPT}
\bibfield{author}{\bibinfo{person}{Charles Packer}, \bibinfo{person}{Vivian Fang}, \bibinfo{person}{Shishir~G. Patil}, \bibinfo{person}{Kevin Lin}, \bibinfo{person}{Sarah Wooders}, {and} \bibinfo{person}{Joseph~E. Gonzalez}.} \bibinfo{year}{2023}\natexlab{}.
\newblock \showarticletitle{MemGPT: Towards LLMs as Operating Systems}.
\newblock \bibinfo{journal}{\emph{CoRR}}  \bibinfo{volume}{abs/2310.08560} (\bibinfo{year}{2023}).
\newblock
\showeprint[arXiv]{2310.08560}
\href{https://doi.org/10.48550/ARXIV.2310.08560}{doi:\nolinkurl{10.48550/ARXIV.2310.08560}}


\bibitem[Phan et~al\mbox{.}(2025)]%
        {hle}
\bibfield{author}{\bibinfo{person}{Long Phan}, \bibinfo{person}{Alice Gatti}, \bibinfo{person}{Ziwen Han}, \bibinfo{person}{Nathaniel Li}, \bibinfo{person}{Josephina Hu}, \bibinfo{person}{Hugh Zhang}, \bibinfo{person}{Sean Shi}, \bibinfo{person}{Michael Choi}, \bibinfo{person}{Anish Agrawal}, \bibinfo{person}{Arnav Chopra}, \bibinfo{person}{Adam Khoja}, \bibinfo{person}{Ryan Kim}, \bibinfo{person}{Jason Hausenloy}, \bibinfo{person}{Oliver Zhang}, \bibinfo{person}{Mantas Mazeika}, \bibinfo{person}{Daron Anderson}, \bibinfo{person}{Tung Nguyen}, \bibinfo{person}{Mobeen Mahmood}, \bibinfo{person}{Fiona Feng}, \bibinfo{person}{Steven~Y. Feng}, \bibinfo{person}{Haoran Zhao}, \bibinfo{person}{Michael Yu}, \bibinfo{person}{Varun Gangal}, \bibinfo{person}{Chelsea Zou}, \bibinfo{person}{Zihan Wang}, \bibinfo{person}{Jessica~P. Wang}, \bibinfo{person}{Pawan Kumar}, \bibinfo{person}{Oleksandr Pokutnyi}, \bibinfo{person}{Robert Gerbicz}, \bibinfo{person}{Serguei Popov}, \bibinfo{person}{John{-}Clark Levin},
  \bibinfo{person}{Mstyslav Kazakov}, \bibinfo{person}{Johannes Schmitt}, \bibinfo{person}{Geoff Galgon}, \bibinfo{person}{Alvaro Sanchez}, \bibinfo{person}{Yongki Lee}, \bibinfo{person}{Will Yeadon}, \bibinfo{person}{Scott Sauers}, \bibinfo{person}{Marc Roth}, \bibinfo{person}{Chidozie Agu}, \bibinfo{person}{S{\o}ren Riis}, \bibinfo{person}{Fabian Giska}, \bibinfo{person}{Saiteja Utpala}, \bibinfo{person}{Zachary Giboney}, \bibinfo{person}{Gashaw~M. Goshu}, \bibinfo{person}{Joan of Arc~Xavier}, \bibinfo{person}{Sarah{-}Jane Crowson}, \bibinfo{person}{Mohinder~Maheshbhai Naiya}, \bibinfo{person}{Noah Burns}, \bibinfo{person}{Lennart Finke}, \bibinfo{person}{Zerui Cheng}, \bibinfo{person}{Hyunwoo Park}, \bibinfo{person}{Francesco Fournier{-}Facio}, \bibinfo{person}{John Wydallis}, \bibinfo{person}{Mark Nandor}, \bibinfo{person}{Ankit Singh}, \bibinfo{person}{Tim Gehrunger}, \bibinfo{person}{Jiaqi Cai}, \bibinfo{person}{Ben McCarty}, \bibinfo{person}{Darling Duclosel}, \bibinfo{person}{Jungbae Nam},
  \bibinfo{person}{Jennifer Zampese}, \bibinfo{person}{Ryan~G. Hoerr}, \bibinfo{person}{Aras Bacho}, \bibinfo{person}{Gautier~Abou Loume}, \bibinfo{person}{Abdallah Galal}, \bibinfo{person}{Hangrui Cao}, \bibinfo{person}{Alexis~C. Garretson}, \bibinfo{person}{Damien Sileo}, \bibinfo{person}{Qiuyu Ren}, \bibinfo{person}{Doru Cojoc}, \bibinfo{person}{Pavel Arkhipov}, \bibinfo{person}{Usman Qazi}, \bibinfo{person}{Lianghui Li}, \bibinfo{person}{Sumeet Motwani}, \bibinfo{person}{Christian~Schr{\"{o}}der de Witt}, \bibinfo{person}{Edwin Taylor}, \bibinfo{person}{Johannes Veith}, \bibinfo{person}{Eric Singer}, \bibinfo{person}{Taylor~D. Hartman}, \bibinfo{person}{Paolo Rissone}, \bibinfo{person}{Jaehyeok Jin}, \bibinfo{person}{Jack Wei~Lun Shi}, \bibinfo{person}{Chris~G. Willcocks}, \bibinfo{person}{Joshua Robinson}, \bibinfo{person}{Aleksandar Mikov}, \bibinfo{person}{Ameya Prabhu}, \bibinfo{person}{Longke Tang}, \bibinfo{person}{Xavier Alapont}, \bibinfo{person}{Justine~Leon Uro}, \bibinfo{person}{Kevin Zhou},
  \bibinfo{person}{Emily de Oliveira~Santos}, \bibinfo{person}{Andrey~Pupasov Maksimov}, \bibinfo{person}{Edward Vendrow}, \bibinfo{person}{Kengo Zenitani}, \bibinfo{person}{Julien Guillod}, \bibinfo{person}{Yuqi Li}, \bibinfo{person}{Joshua Vendrow}, \bibinfo{person}{Vladyslav Kuchkin}, {and} \bibinfo{person}{Ng Ze{-}An}.} \bibinfo{year}{2025}\natexlab{}.
\newblock \showarticletitle{Humanity's Last Exam}.
\newblock \bibinfo{journal}{\emph{CoRR}}  \bibinfo{volume}{abs/2501.14249} (\bibinfo{year}{2025}).
\newblock
\showeprint[arXiv]{2501.14249}
\href{https://doi.org/10.48550/ARXIV.2501.14249}{doi:\nolinkurl{10.48550/ARXIV.2501.14249}}


\bibitem[Qian et~al\mbox{.}(2025)]%
        {MemoRAG}
\bibfield{author}{\bibinfo{person}{Hongjin Qian}, \bibinfo{person}{Zheng Liu}, \bibinfo{person}{Peitian Zhang}, \bibinfo{person}{Kelong Mao}, \bibinfo{person}{Defu Lian}, \bibinfo{person}{Zhicheng Dou}, {and} \bibinfo{person}{Tiejun Huang}.} \bibinfo{year}{2025}\natexlab{}.
\newblock \showarticletitle{MemoRAG: Boosting Long Context Processing with Global Memory-Enhanced Retrieval Augmentation}. In \bibinfo{booktitle}{\emph{Proceedings of the {ACM} on Web Conference 2025, {WWW} 2025, Sydney, NSW, Australia, 28 April 2025- 2 May 2025}}, \bibfield{editor}{\bibinfo{person}{Guodong Long}, \bibinfo{person}{Michale Blumestein}, \bibinfo{person}{Yi~Chang}, \bibinfo{person}{Liane Lewin{-}Eytan}, \bibinfo{person}{Zi~Helen Huang}, {and} \bibinfo{person}{Elad Yom{-}Tov}} (Eds.). \bibinfo{publisher}{{ACM}}, \bibinfo{pages}{2366--2377}.
\newblock
\href{https://doi.org/10.1145/3696410.3714805}{doi:\nolinkurl{10.1145/3696410.3714805}}


\bibitem[Rasmussen et~al\mbox{.}(2025)]%
        {zep}
\bibfield{author}{\bibinfo{person}{Preston Rasmussen}, \bibinfo{person}{Pavlo Paliychuk}, \bibinfo{person}{Travis Beauvais}, \bibinfo{person}{Jack Ryan}, {and} \bibinfo{person}{Daniel Chalef}.} \bibinfo{year}{2025}\natexlab{}.
\newblock \showarticletitle{Zep: {A} Temporal Knowledge Graph Architecture for Agent Memory}.
\newblock \bibinfo{journal}{\emph{CoRR}}  \bibinfo{volume}{abs/2501.13956} (\bibinfo{year}{2025}).
\newblock
\showeprint[arXiv]{2501.13956}
\href{https://doi.org/10.48550/ARXIV.2501.13956}{doi:\nolinkurl{10.48550/ARXIV.2501.13956}}


\bibitem[Schlatt et~al\mbox{.}(2025)]%
        {DistiLLM}
\bibfield{author}{\bibinfo{person}{Ferdinand Schlatt}, \bibinfo{person}{Maik Fr{\"{o}}be}, \bibinfo{person}{Harrisen Scells}, \bibinfo{person}{Shengyao Zhuang}, \bibinfo{person}{Bevan Koopman}, \bibinfo{person}{Guido Zuccon}, \bibinfo{person}{Benno Stein}, \bibinfo{person}{Martin Potthast}, {and} \bibinfo{person}{Matthias Hagen}.} \bibinfo{year}{2025}\natexlab{}.
\newblock \showarticletitle{Rank-DistiLLM: Closing the Effectiveness Gap Between Cross-Encoders and LLMs for Passage Re-ranking}. In \bibinfo{booktitle}{\emph{Advances in Information Retrieval - 47th European Conference on Information Retrieval, {ECIR} 2025, Lucca, Italy, April 6-10, 2025, Proceedings, Part {III}}} \emph{(\bibinfo{series}{Lecture Notes in Computer Science}, Vol.~\bibinfo{volume}{15574})}, \bibfield{editor}{\bibinfo{person}{Claudia Hauff}, \bibinfo{person}{Craig Macdonald}, \bibinfo{person}{Dietmar Jannach}, \bibinfo{person}{Gabriella Kazai}, \bibinfo{person}{Franco~Maria Nardini}, \bibinfo{person}{Fabio Pinelli}, \bibinfo{person}{Fabrizio Silvestri}, {and} \bibinfo{person}{Nicola Tonellotto}} (Eds.). \bibinfo{publisher}{Springer}, \bibinfo{pages}{323--334}.
\newblock
\href{https://doi.org/10.1007/978-3-031-88714-7\_31}{doi:\nolinkurl{10.1007/978-3-031-88714-7\_31}}


\bibitem[Shao et~al\mbox{.}(2023)]%
        {CharacterLLM}
\bibfield{author}{\bibinfo{person}{Yunfan Shao}, \bibinfo{person}{Linyang Li}, \bibinfo{person}{Junqi Dai}, {and} \bibinfo{person}{Xipeng Qiu}.} \bibinfo{year}{2023}\natexlab{}.
\newblock \showarticletitle{Character-LLM: {A} Trainable Agent for Role-Playing}. In \bibinfo{booktitle}{\emph{Proceedings of the 2023 Conference on Empirical Methods in Natural Language Processing, {EMNLP} 2023, Singapore, December 6-10, 2023}}, \bibfield{editor}{\bibinfo{person}{Houda Bouamor}, \bibinfo{person}{Juan Pino}, {and} \bibinfo{person}{Kalika Bali}} (Eds.). \bibinfo{publisher}{Association for Computational Linguistics}, \bibinfo{pages}{13153--13187}.
\newblock
\href{https://doi.org/10.18653/V1/2023.EMNLP-MAIN.814}{doi:\nolinkurl{10.18653/V1/2023.EMNLP-MAIN.814}}


\bibitem[Shao et~al\mbox{.}(2025)]%
        {grpo}
\bibfield{author}{\bibinfo{person}{Zhihong Shao}, \bibinfo{person}{Yuxiang Luo}, \bibinfo{person}{Chengda Lu}, \bibinfo{person}{Z.~Z. Ren}, \bibinfo{person}{Jiewen Hu}, \bibinfo{person}{Tian Ye}, \bibinfo{person}{Zhibin Gou}, \bibinfo{person}{Shirong Ma}, {and} \bibinfo{person}{Xiaokang Zhang}.} \bibinfo{year}{2025}\natexlab{}.
\newblock \showarticletitle{DeepSeekMath-V2: Towards Self-Verifiable Mathematical Reasoning}.
\newblock \bibinfo{journal}{\emph{CoRR}}  \bibinfo{volume}{abs/2511.22570} (\bibinfo{year}{2025}).
\newblock
\showeprint[arXiv]{2511.22570}
\href{https://doi.org/10.48550/ARXIV.2511.22570}{doi:\nolinkurl{10.48550/ARXIV.2511.22570}}


\bibitem[Shi et~al\mbox{.}(2024)]%
        {replug}
\bibfield{author}{\bibinfo{person}{Weijia Shi}, \bibinfo{person}{Sewon Min}, \bibinfo{person}{Michihiro Yasunaga}, \bibinfo{person}{Minjoon Seo}, \bibinfo{person}{Richard James}, \bibinfo{person}{Mike Lewis}, \bibinfo{person}{Luke Zettlemoyer}, {and} \bibinfo{person}{Wen{-}tau Yih}.} \bibinfo{year}{2024}\natexlab{}.
\newblock \showarticletitle{{REPLUG:} Retrieval-Augmented Black-Box Language Models}. In \bibinfo{booktitle}{\emph{Proceedings of the 2024 Conference of the North American Chapter of the Association for Computational Linguistics: Human Language Technologies (Volume 1: Long Papers), {NAACL} 2024, Mexico City, Mexico, June 16-21, 2024}}, \bibfield{editor}{\bibinfo{person}{Kevin Duh}, \bibinfo{person}{Helena G{\'{o}}mez{-}Adorno}, {and} \bibinfo{person}{Steven Bethard}} (Eds.). \bibinfo{publisher}{Association for Computational Linguistics}, \bibinfo{pages}{8371--8384}.
\newblock
\href{https://doi.org/10.18653/V1/2024.NAACL-LONG.463}{doi:\nolinkurl{10.18653/V1/2024.NAACL-LONG.463}}


\bibitem[Sun and Zeng(2025)]%
        {h-mem}
\bibfield{author}{\bibinfo{person}{Haoran Sun} {and} \bibinfo{person}{Shaoning Zeng}.} \bibinfo{year}{2025}\natexlab{}.
\newblock \showarticletitle{Hierarchical Memory for High-Efficiency Long-Term Reasoning in {LLM} Agents}.
\newblock \bibinfo{journal}{\emph{CoRR}}  \bibinfo{volume}{abs/2507.22925} (\bibinfo{year}{2025}).
\newblock
\showeprint[arXiv]{2507.22925}
\href{https://doi.org/10.48550/ARXIV.2507.22925}{doi:\nolinkurl{10.48550/ARXIV.2507.22925}}


\bibitem[Tang et~al\mbox{.}(2025)]%
        {RazorAttention}
\bibfield{author}{\bibinfo{person}{Hanlin Tang}, \bibinfo{person}{Yang Lin}, \bibinfo{person}{Jing Lin}, \bibinfo{person}{Qingsen Han}, \bibinfo{person}{Danning Ke}, \bibinfo{person}{Shikuan Hong}, \bibinfo{person}{Yiwu Yao}, {and} \bibinfo{person}{Gongyi Wang}.} \bibinfo{year}{2025}\natexlab{}.
\newblock \showarticletitle{RazorAttention: Efficient {KV} Cache Compression Through Retrieval Heads}. In \bibinfo{booktitle}{\emph{The Thirteenth International Conference on Learning Representations, {ICLR} 2025, Singapore, April 24-28, 2025}}. \bibinfo{publisher}{OpenReview.net}.
\newblock
\urldef\tempurl%
\url{https://openreview.net/forum?id=tkiZQlL04w}
\showURL{%
\tempurl}


\bibitem[Tang et~al\mbox{.}(2024)]%
        {FoundintheMiddle}
\bibfield{author}{\bibinfo{person}{Raphael Tang}, \bibinfo{person}{Xinyu Zhang}, \bibinfo{person}{Xueguang Ma}, \bibinfo{person}{Jimmy Lin}, {and} \bibinfo{person}{Ferhan Ture}.} \bibinfo{year}{2024}\natexlab{}.
\newblock \showarticletitle{Found in the Middle: Permutation Self-Consistency Improves Listwise Ranking in Large Language Models}. In \bibinfo{booktitle}{\emph{Proceedings of the 2024 Conference of the North American Chapter of the Association for Computational Linguistics: Human Language Technologies (Volume 1: Long Papers), {NAACL} 2024, Mexico City, Mexico, June 16-21, 2024}}, \bibfield{editor}{\bibinfo{person}{Kevin Duh}, \bibinfo{person}{Helena G{\'{o}}mez{-}Adorno}, {and} \bibinfo{person}{Steven Bethard}} (Eds.). \bibinfo{publisher}{Association for Computational Linguistics}, \bibinfo{pages}{2327--2340}.
\newblock
\href{https://doi.org/10.18653/V1/2024.NAACL-LONG.129}{doi:\nolinkurl{10.18653/V1/2024.NAACL-LONG.129}}


\bibitem[Team(2025)]%
        {qwen3technicalreport}
\bibfield{author}{\bibinfo{person}{Qwen Team}.} \bibinfo{year}{2025}\natexlab{}.
\newblock \bibinfo{title}{Qwen3 Technical Report}.
\newblock
\showeprint[arxiv]{2505.09388}~[cs.CL]
\urldef\tempurl%
\url{https://arxiv.org/abs/2505.09388}
\showURL{%
\tempurl}


\bibitem[Vera et~al\mbox{.}(2025)]%
        {EmbeddingGemma}
\bibfield{author}{\bibinfo{person}{Henrique~Schechter Vera}, \bibinfo{person}{Sahil Dua}, \bibinfo{person}{Biao Zhang}, \bibinfo{person}{Daniel Salz}, \bibinfo{person}{Ryan Mullins}, \bibinfo{person}{Sindhu~Raghuram Panyam}, \bibinfo{person}{Sara Smoot}, \bibinfo{person}{Iftekhar Naim}, \bibinfo{person}{Joe Zou}, \bibinfo{person}{Feiyang Chen}, \bibinfo{person}{Daniel Cer}, \bibinfo{person}{Alice Lisak}, \bibinfo{person}{Min Choi}, \bibinfo{person}{Lucas Gonzalez}, \bibinfo{person}{Omar Sanseviero}, \bibinfo{person}{Glenn Cameron}, \bibinfo{person}{Ian Ballantyne}, \bibinfo{person}{Kat Black}, \bibinfo{person}{Kaifeng Chen}, \bibinfo{person}{Weiyi Wang}, \bibinfo{person}{Zhe Li}, \bibinfo{person}{Gus Martins}, \bibinfo{person}{Jinhyuk Lee}, \bibinfo{person}{Mark Sherwood}, \bibinfo{person}{Ju{-}yeong Ji}, \bibinfo{person}{Renjie Wu}, \bibinfo{person}{Jingxiao Zheng}, \bibinfo{person}{Jyotinder Singh}, \bibinfo{person}{Abheesht Sharma}, \bibinfo{person}{Divyashree Sreepathihalli}, \bibinfo{person}{Aashi Jain},
  \bibinfo{person}{Adham Elarabawy}, \bibinfo{person}{AJ Co}, \bibinfo{person}{Andreas Doumanoglou}, \bibinfo{person}{Babak Samari}, \bibinfo{person}{Ben Hora}, \bibinfo{person}{Brian Potetz}, \bibinfo{person}{Dahun Kim}, \bibinfo{person}{Enrique Alfonseca}, \bibinfo{person}{Fedor Moiseev}, \bibinfo{person}{Feng Han}, \bibinfo{person}{Frank~Palma Gomez}, \bibinfo{person}{Gustavo~Hern{\'{a}}ndez {\'{A}}brego}, \bibinfo{person}{Hesen Zhang}, \bibinfo{person}{Hui Hui}, \bibinfo{person}{Jay Han}, \bibinfo{person}{Karan Gill}, \bibinfo{person}{Ke Chen}, \bibinfo{person}{Koert Chen}, \bibinfo{person}{Madhuri Shanbhogue}, \bibinfo{person}{Michael Boratko}, \bibinfo{person}{Paul Suganthan}, \bibinfo{person}{Sai Meher~Karthik Duddu}, \bibinfo{person}{Sandeep Mariserla}, \bibinfo{person}{Setareh Ariafar}, \bibinfo{person}{Shanfeng Zhang}, \bibinfo{person}{Shijie Zhang}, \bibinfo{person}{Simon Baumgartner}, \bibinfo{person}{Sonam Goenka}, \bibinfo{person}{Steve Qiu}, \bibinfo{person}{Tanmaya Dabral},
  \bibinfo{person}{Trevor Walker}, \bibinfo{person}{Vikram Rao}, \bibinfo{person}{Waleed Khawaja}, \bibinfo{person}{Wenlei Zhou}, \bibinfo{person}{Xiaoqi Ren}, \bibinfo{person}{Ye Xia}, \bibinfo{person}{Yichang Chen}, \bibinfo{person}{Yi{-}Ting Chen}, \bibinfo{person}{Zhe Dong}, \bibinfo{person}{Zhongli Ding}, \bibinfo{person}{Francesco Visin}, \bibinfo{person}{Ga{\"{e}}l Liu}, \bibinfo{person}{Jiageng Zhang}, \bibinfo{person}{Kathleen Kenealy}, \bibinfo{person}{Michelle Casbon}, \bibinfo{person}{Ravin Kumar}, \bibinfo{person}{Thomas Mesnard}, \bibinfo{person}{Zach Gleicher}, \bibinfo{person}{Cormac Brick}, \bibinfo{person}{Olivier Lacombe}, \bibinfo{person}{Adam Roberts}, \bibinfo{person}{Qin Yin}, \bibinfo{person}{Yun{-}Hsuan Sung}, \bibinfo{person}{Raphael Hoffmann}, \bibinfo{person}{Tris Warkentin}, \bibinfo{person}{Armand Joulin}, \bibinfo{person}{Tom Duerig}, {and} \bibinfo{person}{Mojtaba Seyedhosseini}.} \bibinfo{year}{2025}\natexlab{}.
\newblock \showarticletitle{EmbeddingGemma: Powerful and Lightweight Text Representations}.
\newblock \bibinfo{journal}{\emph{CoRR}}  \bibinfo{volume}{abs/2509.20354} (\bibinfo{year}{2025}).
\newblock
\showeprint[arXiv]{2509.20354}
\href{https://doi.org/10.48550/ARXIV.2509.20354}{doi:\nolinkurl{10.48550/ARXIV.2509.20354}}


\bibitem[Vijay et~al\mbox{.}(2025)]%
        {think-before-retrive}
\bibfield{author}{\bibinfo{person}{Supriti Vijay}, \bibinfo{person}{Aman Priyanshu}, \bibinfo{person}{Anu Vellore}, \bibinfo{person}{Baturay Saglam}, {and} \bibinfo{person}{Amin Karbasi}.} \bibinfo{year}{2025}\natexlab{}.
\newblock \bibinfo{title}{Think Before You Retrieve: Learning Test-Time Adaptive Search with Small Language Models}.
\newblock
\showeprint[arxiv]{2511.07581}~[cs.AI]
\urldef\tempurl%
\url{https://arxiv.org/abs/2511.07581}
\showURL{%
\tempurl}


\bibitem[Wang et~al\mbox{.}(2023)]%
        {simlm}
\bibfield{author}{\bibinfo{person}{Liang Wang}, \bibinfo{person}{Nan Yang}, \bibinfo{person}{Xiaolong Huang}, \bibinfo{person}{Binxing Jiao}, \bibinfo{person}{Linjun Yang}, \bibinfo{person}{Daxin Jiang}, \bibinfo{person}{Rangan Majumder}, {and} \bibinfo{person}{Furu Wei}.} \bibinfo{year}{2023}\natexlab{}.
\newblock \showarticletitle{SimLM: Pre-training with Representation Bottleneck for Dense Passage Retrieval}. In \bibinfo{booktitle}{\emph{Proceedings of the 61st Annual Meeting of the Association for Computational Linguistics (Volume 1: Long Papers), {ACL} 2023, Toronto, Canada, July 9-14, 2023}}, \bibfield{editor}{\bibinfo{person}{Anna Rogers}, \bibinfo{person}{Jordan~L. Boyd{-}Graber}, {and} \bibinfo{person}{Naoaki Okazaki}} (Eds.). \bibinfo{publisher}{Association for Computational Linguistics}, \bibinfo{pages}{2244--2258}.
\newblock
\href{https://doi.org/10.18653/V1/2023.ACL-LONG.125}{doi:\nolinkurl{10.18653/V1/2023.ACL-LONG.125}}


\bibitem[Wang et~al\mbox{.}(2024)]%
        {memoryllm}
\bibfield{author}{\bibinfo{person}{Yu Wang}, \bibinfo{person}{Yifan Gao}, \bibinfo{person}{Xiusi Chen}, \bibinfo{person}{Haoming Jiang}, \bibinfo{person}{Shiyang Li}, \bibinfo{person}{Jingfeng Yang}, \bibinfo{person}{Qingyu Yin}, \bibinfo{person}{Zheng Li}, \bibinfo{person}{Xian Li}, \bibinfo{person}{Bing Yin}, \bibinfo{person}{Jingbo Shang}, {and} \bibinfo{person}{Julian~J. McAuley}.} \bibinfo{year}{2024}\natexlab{}.
\newblock \showarticletitle{{MEMORYLLM:} Towards Self-Updatable Large Language Models}. In \bibinfo{booktitle}{\emph{Forty-first International Conference on Machine Learning, {ICML} 2024, Vienna, Austria, July 21-27, 2024}}. \bibinfo{publisher}{OpenReview.net}.
\newblock
\urldef\tempurl%
\url{https://openreview.net/forum?id=p0lKWzdikQ}
\showURL{%
\tempurl}


\bibitem[Wei et~al\mbox{.}(2025)]%
        {Wei2025MLPMA}
\bibfield{author}{\bibinfo{person}{Rubin Wei}, \bibinfo{person}{Jiaqi Cao}, \bibinfo{person}{Jiarui Wang}, \bibinfo{person}{Jushi Kai}, \bibinfo{person}{Qipeng Guo}, \bibinfo{person}{Bowen Zhou}, {and} \bibinfo{person}{Zhouhan Lin}.} \bibinfo{year}{2025}\natexlab{}.
\newblock \showarticletitle{MLP Memory: A Retriever-Pretrained Memory for Large Language Models}.
\newblock
\urldef\tempurl%
\url{https://api.semanticscholar.org/CorpusID:281658735}
\showURL{%
\tempurl}


\bibitem[Wisznia et~al\mbox{.}(2025)]%
        {AreOptimalAlg}
\bibfield{author}{\bibinfo{person}{Juan Wisznia}, \bibinfo{person}{Cecilia Bola{\~{n}}os}, \bibinfo{person}{Juan Tollo}, \bibinfo{person}{Giovanni Marraffini}, \bibinfo{person}{Agust{\'{\i}}n Gianolini}, \bibinfo{person}{Noe Hsueh}, {and} \bibinfo{person}{Luciano~Del Corro}.} \bibinfo{year}{2025}\natexlab{}.
\newblock \showarticletitle{Are Optimal Algorithms Still Optimal? Rethinking Sorting in LLM-Based Pairwise Ranking with Batching and Caching}. In \bibinfo{booktitle}{\emph{Proceedings of the 63rd Annual Meeting of the Association for Computational Linguistics (Volume 2: Short Papers), {ACL} 2025, Vienna, Austria, July 27 - August 1, 2025}}, \bibfield{editor}{\bibinfo{person}{Wanxiang Che}, \bibinfo{person}{Joyce Nabende}, \bibinfo{person}{Ekaterina Shutova}, {and} \bibinfo{person}{Mohammad~Taher Pilehvar}} (Eds.). \bibinfo{publisher}{Association for Computational Linguistics}, \bibinfo{pages}{1064--1072}.
\newblock
\href{https://doi.org/10.18653/V1/2025.ACL-SHORT.83}{doi:\nolinkurl{10.18653/V1/2025.ACL-SHORT.83}}


\bibitem[Wu et~al\mbox{.}(2025b)]%
        {longmemeval}
\bibfield{author}{\bibinfo{person}{Di Wu}, \bibinfo{person}{Hongwei Wang}, \bibinfo{person}{Wenhao Yu}, \bibinfo{person}{Yuwei Zhang}, \bibinfo{person}{Kai{-}Wei Chang}, {and} \bibinfo{person}{Dong Yu}.} \bibinfo{year}{2025}\natexlab{b}.
\newblock \showarticletitle{LongMemEval: Benchmarking Chat Assistants on Long-Term Interactive Memory}. In \bibinfo{booktitle}{\emph{The Thirteenth International Conference on Learning Representations, {ICLR} 2025, Singapore, April 24-28, 2025}}. \bibinfo{publisher}{OpenReview.net}.
\newblock
\urldef\tempurl%
\url{https://openreview.net/forum?id=pZiyCaVuti}
\showURL{%
\tempurl}


\bibitem[Wu et~al\mbox{.}(2025a)]%
        {webdancer}
\bibfield{author}{\bibinfo{person}{Jialong Wu}, \bibinfo{person}{Baixuan Li}, \bibinfo{person}{Runnan Fang}, \bibinfo{person}{Wenbiao Yin}, \bibinfo{person}{Liwen Zhang}, \bibinfo{person}{Zhengwei Tao}, \bibinfo{person}{Dingchu Zhang}, \bibinfo{person}{Zekun Xi}, \bibinfo{person}{Yong Jiang}, \bibinfo{person}{Pengjun Xie}, \bibinfo{person}{Fei Huang}, {and} \bibinfo{person}{Jingren Zhou}.} \bibinfo{year}{2025}\natexlab{a}.
\newblock \showarticletitle{WebDancer: Towards Autonomous Information Seeking Agency}.
\newblock \bibinfo{journal}{\emph{CoRR}}  \bibinfo{volume}{abs/2505.22648} (\bibinfo{year}{2025}).
\newblock
\showeprint[arXiv]{2505.22648}
\href{https://doi.org/10.48550/ARXIV.2505.22648}{doi:\nolinkurl{10.48550/ARXIV.2505.22648}}


\bibitem[Wu et~al\mbox{.}(2025c)]%
        {webwalker}
\bibfield{author}{\bibinfo{person}{Jialong Wu}, \bibinfo{person}{Wenbiao Yin}, \bibinfo{person}{Yong Jiang}, \bibinfo{person}{Zhenglin Wang}, \bibinfo{person}{Zekun Xi}, \bibinfo{person}{Runnan Fang}, \bibinfo{person}{Linhai Zhang}, \bibinfo{person}{Yulan He}, \bibinfo{person}{Deyu Zhou}, \bibinfo{person}{Pengjun Xie}, {and} \bibinfo{person}{Fei Huang}.} \bibinfo{year}{2025}\natexlab{c}.
\newblock \showarticletitle{WebWalker: Benchmarking LLMs in Web Traversal}. In \bibinfo{booktitle}{\emph{Proceedings of the 63rd Annual Meeting of the Association for Computational Linguistics (Volume 1: Long Papers), {ACL} 2025, Vienna, Austria, July 27 - August 1, 2025}}, \bibfield{editor}{\bibinfo{person}{Wanxiang Che}, \bibinfo{person}{Joyce Nabende}, \bibinfo{person}{Ekaterina Shutova}, {and} \bibinfo{person}{Mohammad~Taher Pilehvar}} (Eds.). \bibinfo{publisher}{Association for Computational Linguistics}, \bibinfo{pages}{10290--10305}.
\newblock
\urldef\tempurl%
\url{https://aclanthology.org/2025.acl-long.508/}
\showURL{%
\tempurl}


\bibitem[Xu et~al\mbox{.}(2025b)]%
        {amem}
\bibfield{author}{\bibinfo{person}{Wujiang Xu}, \bibinfo{person}{Zujie Liang}, \bibinfo{person}{Kai Mei}, \bibinfo{person}{Hang Gao}, \bibinfo{person}{Juntao Tan}, {and} \bibinfo{person}{Yongfeng Zhang}.} \bibinfo{year}{2025}\natexlab{b}.
\newblock \showarticletitle{{A-MEM:} Agentic Memory for {LLM} Agents}.
\newblock \bibinfo{journal}{\emph{CoRR}}  \bibinfo{volume}{abs/2502.12110} (\bibinfo{year}{2025}).
\newblock
\showeprint[arXiv]{2502.12110}
\href{https://doi.org/10.48550/ARXIV.2502.12110}{doi:\nolinkurl{10.48550/ARXIV.2502.12110}}


\bibitem[Xu et~al\mbox{.}(2025a)]%
        {scarlet}
\bibfield{author}{\bibinfo{person}{Yilong Xu}, \bibinfo{person}{Jinhua Gao}, \bibinfo{person}{Xiaoming Yu}, \bibinfo{person}{Yuanhai Xue}, \bibinfo{person}{Baolong Bi}, \bibinfo{person}{Huawei Shen}, {and} \bibinfo{person}{Xueqi Cheng}.} \bibinfo{year}{2025}\natexlab{a}.
\newblock \showarticletitle{Training a Utility-based Retriever Through Shared Context Attribution for Retrieval-Augmented Language Models}. In \bibinfo{booktitle}{\emph{Proceedings of the 2025 Conference on Empirical Methods in Natural Language Processing, {EMNLP} 2025, Suzhou, China, November 4-9, 2025}}, \bibfield{editor}{\bibinfo{person}{Christos Christodoulopoulos}, \bibinfo{person}{Tanmoy Chakraborty}, \bibinfo{person}{Carolyn Rose}, {and} \bibinfo{person}{Violet Peng}} (Eds.). \bibinfo{publisher}{Association for Computational Linguistics}, \bibinfo{pages}{629--648}.
\newblock
\href{https://doi.org/10.18653/V1/2025.EMNLP-MAIN.33}{doi:\nolinkurl{10.18653/V1/2025.EMNLP-MAIN.33}}


\bibitem[Yan et~al\mbox{.}(2025)]%
        {O1Embedder}
\bibfield{author}{\bibinfo{person}{Ruiran Yan}, \bibinfo{person}{Zheng Liu}, {and} \bibinfo{person}{Defu Lian}.} \bibinfo{year}{2025}\natexlab{}.
\newblock \showarticletitle{{O1} Embedder: Let Retrievers Think Before Action}.
\newblock \bibinfo{journal}{\emph{CoRR}}  \bibinfo{volume}{abs/2502.07555} (\bibinfo{year}{2025}).
\newblock
\showeprint[arXiv]{2502.07555}
\href{https://doi.org/10.48550/ARXIV.2502.07555}{doi:\nolinkurl{10.48550/ARXIV.2502.07555}}


\bibitem[Yang et~al\mbox{.}(2018)]%
        {hotpot-qa}
\bibfield{author}{\bibinfo{person}{Zhilin Yang}, \bibinfo{person}{Peng Qi}, \bibinfo{person}{Saizheng Zhang}, \bibinfo{person}{Yoshua Bengio}, \bibinfo{person}{William~W. Cohen}, \bibinfo{person}{Ruslan Salakhutdinov}, {and} \bibinfo{person}{Christopher~D. Manning}.} \bibinfo{year}{2018}\natexlab{}.
\newblock \showarticletitle{HotpotQA: {A} Dataset for Diverse, Explainable Multi-hop Question Answering}. In \bibinfo{booktitle}{\emph{Proceedings of the 2018 Conference on Empirical Methods in Natural Language Processing, Brussels, Belgium, October 31 - November 4, 2018}}, \bibfield{editor}{\bibinfo{person}{Ellen Riloff}, \bibinfo{person}{David Chiang}, \bibinfo{person}{Julia Hockenmaier}, {and} \bibinfo{person}{Jun'ichi Tsujii}} (Eds.). \bibinfo{publisher}{Association for Computational Linguistics}, \bibinfo{pages}{2369--2380}.
\newblock
\href{https://doi.org/10.18653/V1/D18-1259}{doi:\nolinkurl{10.18653/V1/D18-1259}}


\bibitem[Yu et~al\mbox{.}(2025a)]%
        {memagent}
\bibfield{author}{\bibinfo{person}{Hongli Yu}, \bibinfo{person}{Tinghong Chen}, \bibinfo{person}{Jiangtao Feng}, \bibinfo{person}{Jiangjie Chen}, \bibinfo{person}{Weinan Dai}, \bibinfo{person}{Qiying Yu}, \bibinfo{person}{Ya{-}Qin Zhang}, \bibinfo{person}{Wei{-}Ying Ma}, \bibinfo{person}{Jingjing Liu}, \bibinfo{person}{Mingxuan Wang}, {and} \bibinfo{person}{Hao Zhou}.} \bibinfo{year}{2025}\natexlab{a}.
\newblock \showarticletitle{MemAgent: Reshaping Long-Context {LLM} with Multi-Conv RL-based Memory Agent}.
\newblock \bibinfo{journal}{\emph{CoRR}}  \bibinfo{volume}{abs/2507.02259} (\bibinfo{year}{2025}).
\newblock
\showeprint[arXiv]{2507.02259}
\href{https://doi.org/10.48550/ARXIV.2507.02259}{doi:\nolinkurl{10.48550/ARXIV.2507.02259}}


\bibitem[Yu et~al\mbox{.}(2025b)]%
        {dapo}
\bibfield{author}{\bibinfo{person}{Qiying Yu}, \bibinfo{person}{Zheng Zhang}, \bibinfo{person}{Ruofei Zhu}, \bibinfo{person}{Yufeng Yuan}, \bibinfo{person}{Xiaochen Zuo}, \bibinfo{person}{Yu Yue}, \bibinfo{person}{Tiantian Fan}, \bibinfo{person}{Gaohong Liu}, \bibinfo{person}{Lingjun Liu}, \bibinfo{person}{Xin Liu}, \bibinfo{person}{Haibin Lin}, \bibinfo{person}{Zhiqi Lin}, \bibinfo{person}{Bole Ma}, \bibinfo{person}{Guangming Sheng}, \bibinfo{person}{Yuxuan Tong}, \bibinfo{person}{Chi Zhang}, \bibinfo{person}{Mofan Zhang}, \bibinfo{person}{Wang Zhang}, \bibinfo{person}{Hang Zhu}, \bibinfo{person}{Jinhua Zhu}, \bibinfo{person}{Jiaze Chen}, \bibinfo{person}{Jiangjie Chen}, \bibinfo{person}{Chengyi Wang}, \bibinfo{person}{Hongli Yu}, \bibinfo{person}{Weinan Dai}, \bibinfo{person}{Yuxuan Song}, \bibinfo{person}{Xiangpeng Wei}, \bibinfo{person}{Hao Zhou}, \bibinfo{person}{Jingjing Liu}, \bibinfo{person}{Wei{-}Ying Ma}, \bibinfo{person}{Ya{-}Qin Zhang}, \bibinfo{person}{Lin Yan}, \bibinfo{person}{Mu Qiao},
  \bibinfo{person}{Yonghui Wu}, {and} \bibinfo{person}{Mingxuan Wang}.} \bibinfo{year}{2025}\natexlab{b}.
\newblock \showarticletitle{{DAPO:} An Open-Source {LLM} Reinforcement Learning System at Scale}.
\newblock \bibinfo{journal}{\emph{CoRR}}  \bibinfo{volume}{abs/2503.14476} (\bibinfo{year}{2025}).
\newblock
\showeprint[arXiv]{2503.14476}
\href{https://doi.org/10.48550/ARXIV.2503.14476}{doi:\nolinkurl{10.48550/ARXIV.2503.14476}}


\bibitem[Zamani and Bendersky(2024)]%
        {stochastic-rag}
\bibfield{author}{\bibinfo{person}{Hamed Zamani} {and} \bibinfo{person}{Michael Bendersky}.} \bibinfo{year}{2024}\natexlab{}.
\newblock \showarticletitle{Stochastic {RAG:} End-to-End Retrieval-Augmented Generation through Expected Utility Maximization}. In \bibinfo{booktitle}{\emph{Proceedings of the 47th International {ACM} {SIGIR} Conference on Research and Development in Information Retrieval, {SIGIR} 2024, Washington DC, USA, July 14-18, 2024}}, \bibfield{editor}{\bibinfo{person}{Grace~Hui Yang}, \bibinfo{person}{Hongning Wang}, \bibinfo{person}{Sam Han}, \bibinfo{person}{Claudia Hauff}, \bibinfo{person}{Guido Zuccon}, {and} \bibinfo{person}{Yi~Zhang}} (Eds.). \bibinfo{publisher}{{ACM}}, \bibinfo{pages}{2641--2646}.
\newblock
\href{https://doi.org/10.1145/3626772.3657923}{doi:\nolinkurl{10.1145/3626772.3657923}}


\bibitem[Zhang et~al\mbox{.}(2025a)]%
        {realm}
\bibfield{author}{\bibinfo{person}{Hengran Zhang}, \bibinfo{person}{Keping Bi}, \bibinfo{person}{Jiafeng Guo}, \bibinfo{person}{Jiaming Zhang}, \bibinfo{person}{Shuaiqiang Wang}, \bibinfo{person}{Dawei Yin}, {and} \bibinfo{person}{Xueqi Cheng}.} \bibinfo{year}{2025}\natexlab{a}.
\newblock \showarticletitle{LLM-Specific Utility: {A} New Perspective for Retrieval-Augmented Generation}.
\newblock \bibinfo{journal}{\emph{CoRR}}  \bibinfo{volume}{abs/2510.11358} (\bibinfo{year}{2025}).
\newblock
\showeprint[arXiv]{2510.11358}
\href{https://doi.org/10.48550/ARXIV.2510.11358}{doi:\nolinkurl{10.48550/ARXIV.2510.11358}}


\bibitem[Zhang et~al\mbox{.}(2025b)]%
        {llm-specific-utility}
\bibfield{author}{\bibinfo{person}{Hengran Zhang}, \bibinfo{person}{Keping Bi}, \bibinfo{person}{Jiafeng Guo}, \bibinfo{person}{Jiaming Zhang}, \bibinfo{person}{Shuaiqiang Wang}, \bibinfo{person}{Dawei Yin}, {and} \bibinfo{person}{Xueqi Cheng}.} \bibinfo{year}{2025}\natexlab{b}.
\newblock \showarticletitle{LLM-Specific Utility: {A} New Perspective for Retrieval-Augmented Generation}.
\newblock \bibinfo{journal}{\emph{CoRR}}  \bibinfo{volume}{abs/2510.11358} (\bibinfo{year}{2025}).
\newblock
\showeprint[arXiv]{2510.11358}
\href{https://doi.org/10.48550/ARXIV.2510.11358}{doi:\nolinkurl{10.48550/ARXIV.2510.11358}}


\bibitem[Zhang et~al\mbox{.}(2025c)]%
        {rearank}
\bibfield{author}{\bibinfo{person}{Le Zhang}, \bibinfo{person}{Bo Wang}, \bibinfo{person}{Xipeng Qiu}, \bibinfo{person}{Siva Reddy}, {and} \bibinfo{person}{Aishwarya Agrawal}.} \bibinfo{year}{2025}\natexlab{c}.
\newblock \showarticletitle{{REARANK:} Reasoning Re-ranking Agent via Reinforcement Learning}.
\newblock \bibinfo{journal}{\emph{CoRR}}  \bibinfo{volume}{abs/2505.20046} (\bibinfo{year}{2025}).
\newblock
\showeprint[arXiv]{2505.20046}
\href{https://doi.org/10.48550/ARXIV.2505.20046}{doi:\nolinkurl{10.48550/ARXIV.2505.20046}}


\bibitem[Zhong et~al\mbox{.}(2024)]%
        {memory-bank}
\bibfield{author}{\bibinfo{person}{Wanjun Zhong}, \bibinfo{person}{Lianghong Guo}, \bibinfo{person}{Qiqi Gao}, \bibinfo{person}{He Ye}, {and} \bibinfo{person}{Yanlin Wang}.} \bibinfo{year}{2024}\natexlab{}.
\newblock \showarticletitle{MemoryBank: Enhancing Large Language Models with Long-Term Memory}. In \bibinfo{booktitle}{\emph{Thirty-Eighth {AAAI} Conference on Artificial Intelligence, {AAAI} 2024, Thirty-Sixth Conference on Innovative Applications of Artificial Intelligence, {IAAI} 2024, Fourteenth Symposium on Educational Advances in Artificial Intelligence, {EAAI} 2014, February 20-27, 2024, Vancouver, Canada}}, \bibfield{editor}{\bibinfo{person}{Michael~J. Wooldridge}, \bibinfo{person}{Jennifer~G. Dy}, {and} \bibinfo{person}{Sriraam Natarajan}} (Eds.). \bibinfo{publisher}{{AAAI} Press}, \bibinfo{pages}{19724--19731}.
\newblock
\href{https://doi.org/10.1609/AAAI.V38I17.29946}{doi:\nolinkurl{10.1609/AAAI.V38I17.29946}}


\bibitem[Zhou et~al\mbox{.}(2025)]%
        {mem1}
\bibfield{author}{\bibinfo{person}{Zijian Zhou}, \bibinfo{person}{Ao Qu}, \bibinfo{person}{Zhaoxuan Wu}, \bibinfo{person}{Sunghwan Kim}, \bibinfo{person}{Alok Prakash}, \bibinfo{person}{Daniela Rus}, \bibinfo{person}{Jinhua Zhao}, \bibinfo{person}{Bryan Kian~Hsiang Low}, {and} \bibinfo{person}{Paul~Pu Liang}.} \bibinfo{year}{2025}\natexlab{}.
\newblock \showarticletitle{{MEM1:} Learning to Synergize Memory and Reasoning for Efficient Long-Horizon Agents}.
\newblock \bibinfo{journal}{\emph{CoRR}}  \bibinfo{volume}{abs/2506.15841} (\bibinfo{year}{2025}).
\newblock
\showeprint[arXiv]{2506.15841}
\href{https://doi.org/10.48550/ARXIV.2506.15841}{doi:\nolinkurl{10.48550/ARXIV.2506.15841}}


\bibitem[Zhuang et~al\mbox{.}(2023)]%
        {RankT5}
\bibfield{author}{\bibinfo{person}{Honglei Zhuang}, \bibinfo{person}{Zhen Qin}, \bibinfo{person}{Rolf Jagerman}, \bibinfo{person}{Kai Hui}, \bibinfo{person}{Ji Ma}, \bibinfo{person}{Jing Lu}, \bibinfo{person}{Jianmo Ni}, \bibinfo{person}{Xuanhui Wang}, {and} \bibinfo{person}{Michael Bendersky}.} \bibinfo{year}{2023}\natexlab{}.
\newblock \showarticletitle{RankT5: Fine-Tuning {T5} for Text Ranking with Ranking Losses}. In \bibinfo{booktitle}{\emph{Proceedings of the 46th International {ACM} {SIGIR} Conference on Research and Development in Information Retrieval, {SIGIR} 2023, Taipei, Taiwan, July 23-27, 2023}}, \bibfield{editor}{\bibinfo{person}{Hsin{-}Hsi Chen}, \bibinfo{person}{Wei{-}Jou~(Edward) Duh}, \bibinfo{person}{Hen{-}Hsen Huang}, \bibinfo{person}{Makoto~P. Kato}, \bibinfo{person}{Josiane Mothe}, {and} \bibinfo{person}{Barbara Poblete}} (Eds.). \bibinfo{publisher}{{ACM}}, \bibinfo{pages}{2308--2313}.
\newblock
\href{https://doi.org/10.1145/3539618.3592047}{doi:\nolinkurl{10.1145/3539618.3592047}}


\bibitem[Zhuang et~al\mbox{.}(2025)]%
        {RankR1}
\bibfield{author}{\bibinfo{person}{Shengyao Zhuang}, \bibinfo{person}{Xueguang Ma}, \bibinfo{person}{Bevan Koopman}, \bibinfo{person}{Jimmy Lin}, {and} \bibinfo{person}{Guido Zuccon}.} \bibinfo{year}{2025}\natexlab{}.
\newblock \showarticletitle{Rank-R1: Enhancing Reasoning in LLM-based Document Rerankers via Reinforcement Learning}.
\newblock \bibinfo{journal}{\emph{CoRR}}  \bibinfo{volume}{abs/2503.06034} (\bibinfo{year}{2025}).
\newblock
\showeprint[arXiv]{2503.06034}
\href{https://doi.org/10.48550/ARXIV.2503.06034}{doi:\nolinkurl{10.48550/ARXIV.2503.06034}}


\end{thebibliography}

\newpage

\appendix

\section{Experimental Details}
\label{sec:appendix-experimental-details}

\textbf{Implementation details.}
We use Qwen3-4B-Thinking~\cite{qwen3technicalreport} as the base model and train it for three rounds following the training method introduced in Section~\ref{sec:training_strategy}. During training, we adopt a 128k long input window, a 16k formal output window, and a 4k buffer window. The batch size is set to 32, and the number of GRPO~\cite{grpo} sampling times $n$ for each sample is set to 6. The RL training experiments of the memory proxy are conducted on an 8-GPU H200 machine. The working LLM calls the external Qwen3-30B-A3B-Instruct API.

For training data preparation, we sample training data from the training sets of multiple memory-related benchmarks and restructure them to fit our method. We split the sessions and label each with a session ID following the input format specified in Section~\ref{memory-proxy-reasoning}. For excessively long memories (exceeding 128K tokens), we sample memory content while ensuring that all relevant memories are included. More details are in Appendix~\ref{sec:appendix-deep-research}.

\textbf{Session construction.}
When a benchmark provides session boundaries, we keep the original boundaries. For deep research trajectories, one search step or reasoning step is treated as one session. This session-level design is cheaper than turn-level retrieval and keeps the full text inside each selected session. Its limitation is that noisy sessions may still contain irrelevant turns; turn-level or hierarchical retrieval is left as future work.

\textbf{Coarse filtering.}
We use BGE-M3 for the first-stage filter only when the full history is longer than the 128K proxy window. We embed the current query and each session, rank sessions by cosine similarity, and keep top sessions until the token budget is reached. We do not use a fixed similarity threshold. The full text of each kept session is then passed to the proxy.

\textbf{Top-$k$ setting.}
The proxy outputs a fixed top-10 ranked list. We use this fixed list size to keep the comparison stable across datasets and baselines. The working LLM receives the same top-$k$ budget under each method. We also reserve a top-$k$ sensitivity table to report whether smaller or larger $k$ changes the cost-performance trade-off. For fair comparison across datasets, the token cost in this table counts only the memory/session tokens introduced by top-$k$ selection and excludes the current query and fixed task prompt.

\textbf{Cross-working-LLM setting.}
The memory proxy is optimized with task feedback from Qwen3-30B-A3B-Instruct. At test time, the retrieved memory is evaluated with both Qwen3-30B-A3B-Instruct and DeepSeek V3.2 as working LLMs. This setting tests whether the proxy learns a retrieval policy that transfers across working LLMs.

\textbf{Training cost accounting.}
We report training cost using GPU node hours, working LLM calls, input/output tokens, and API cost. All three training rounds are completed on one 8$\times$H200 node. The reward model calls Qwen3-30B-A3B-Instruct through the SiliconFlow API, charged at RMB~0.70 per million input tokens and RMB~2.80 per million output tokens. The API token counts in Table~\ref{tab:training-cost} are billing-level estimates reconstructed from this price schedule and the final provider bill. For top-10 reward evaluation, dense evaluation would require the no-memory baseline plus every rank cutoff, i.e., 11 working-LLM calls. Fibonacci sampling evaluates $\{0,1,2,3,5,8,10\}$ instead, reducing the reward-evaluation calls to 7 per rollout.

\textbf{Direct-rank ablation.}
To test the need for reasoning before retrieval, we add a direct-rank variant. The standard MemSifter proxy follows a Think-and-Rank format, first emitting ``\texttt{<think>}'' and then ``\texttt{<ranking>}''. The direct-rank variant removes the ``\texttt{<think>}'' output requirement and asks the proxy to output only ``\texttt{<ranking>}''. All other settings are kept the same, including the base model, train/test split, top-$k$ budget, working LLM, and evaluation metric. The reporting set covers LoCoMo-32K, LongMemEval-1M, and WebDancer-128K, with F1, output tokens, and latency.

\begin{table}[t]
\small
\centering
\caption{Train/test split statistics. Train and test sets are strictly disjoint.}
\label{tab:dataset-splits}
\setlength{\tabcolsep}{2.5mm}{
\begin{tabular}{lcrr}
\toprule
\textbf{Dataset} & \textbf{Scale} & \textbf{Train} & \textbf{Test} \\
\midrule
LoCoMo & 32K & 1,136 & 400 \\
LongMemEval & 128K & 350 & 150 \\
LongMemEval & 1M & 350 & 150 \\
PersonaMem & 32K & 0 & 400 \\
PersonaMem & 128K & 2,303 & 400 \\
PerM-V2 & 128K & 3,000 & 400 \\
ZH4O & 128K & 668 & 400 \\
HotpotQA & 128K & 3,000 & 400 \\
WebWalker & 128K & 3,000 & 400 \\
WebDancer & 128K & 3,000 & 400 \\
\midrule
\textbf{Total} & -- & \textbf{15,807} & \textbf{3,500} \\
\bottomrule
\end{tabular}
}
\end{table}

\begin{table}[t]
\small
\centering
\caption{Coarse filtering statistics.}
\label{tab:coarse-filtering-stats}
\setlength{\tabcolsep}{1.1mm}{
\begin{tabular}{lcc}
\toprule
\textbf{Dataset} & \textbf{Recall} & \textbf{Ret. Tok.} \\
\midrule
LoCoMo-32K & 100.00 & 22.2K \\
LongMemEval-128K & 100.00 & 105.0K \\
LongMemEval-1M & 94.88 & 128.0K \\
PersonaMem-32K & 100.00 & 24.3K \\
PersonaMem-128K & 100.00 & 117.1K \\
PerM-V2-128K & 96.50 & 128.0K \\
ZH4O-128K & 100.00 & 20.0K \\
\bottomrule
\end{tabular}
}
\end{table}

\begin{table}[t]
\small
\centering
\caption{Training cost accounting.}
\label{tab:training-cost}
\setlength{\tabcolsep}{2.0mm}{
\begin{tabular}{p{0.42\linewidth}p{0.44\linewidth}}
\toprule
\textbf{Item} & \textbf{Value} \\
\midrule
GPU hardware & 1 node, 8$\times$H200 \\
GPU node hours & $\approx$96 node-hours \\
Training rounds & 3 \\
Working LLM calls & $\approx$200K \\
Input tokens & $\approx$250M \\
Output tokens & $\approx$42M \\
API cost & $\approx$RMB~292 (\$40) \\
Fibonacci sampling call saving & 7 vs. 11 calls per rollout (36.4\%) \\
\bottomrule
\end{tabular}
}
\end{table}

\begin{table}[t]
\small
\centering
\caption{Inference-time top-$k$ sensitivity results. Tok. denotes the estimated average memory/session tokens introduced by top-$k$ selection, excluding the current query and fixed task prompt.}
\label{tab:topk-sensitivity}
\setlength{\tabcolsep}{1.4mm}{
\begin{tabular}{lrrrrrr}
\toprule
\textbf{Top-$k$} & \multicolumn{2}{c}{\textbf{LoCoMo 32K}} & \multicolumn{2}{c}{\textbf{LongMemEval 1M}} & \multicolumn{2}{c}{\textbf{WebDancer 128K}} \\
\cmidrule(lr){2-3}\cmidrule(lr){4-5}\cmidrule(lr){6-7}
& F1 & Tok. & F1 & Tok. & F1 & Tok. \\
\midrule
3  & 40.02 & 2.4K  & 30.38 & 6.5K  & 36.64 & 6.9K \\
5  & 41.17 & 4.0K  & 31.81 & 10.9K & 37.10 & 11.6K \\
10 & 41.79 & 8.0K  & 33.32 & 21.8K & 38.21 & 23.1K \\
15 & 41.46 & 12.0K & 32.95 & 32.7K & 37.87 & 34.7K \\
\bottomrule
\end{tabular}
}
\end{table}

\begin{table}[t]
\small
\centering
\caption{PersonaMem multiple-choice accuracy. Accuracy is computed by normalizing each generated answer to the predicted option before exact-match scoring. The best and second best of each setting are in \textbf{bold} and \underline{underlined}.}
\label{tab:personamem-accuracy}
\setlength{\tabcolsep}{1.8mm}{
\begin{tabular}{lcccc}
\toprule
\multirow{2}{*}{\textbf{Method}} & \multicolumn{2}{c}{\textbf{DeepSeek V3.2}} & \multicolumn{2}{c}{\textbf{Qwen3-30B}} \\
\cmidrule(lr){2-3}\cmidrule(lr){4-5}
& 32K & 128K & 32K & 128K \\
\midrule
BGE-M3 & 31.50 & 23.75 & 36.00 & 30.50 \\
GemmaEmb & 32.75 & 23.50 & 31.75 & 27.75 \\
Nemori & 32.75 & 25.25 & 34.25 & 29.75 \\
Mem0 & 35.00 & 26.75 & 35.75 & 28.00 \\
MemAgent & 29.50 & 24.75 & \underline{36.75} & 27.75 \\
Mem-$\alpha$ & \underline{37.25} & 25.75 & 34.00 & 28.75 \\
HippoRAG & 37.00 & 24.75 & 33.25 & 27.25 \\
A-MEM & 33.00 & 22.75 & 35.25 & 27.25 \\
Rearank & 34.50 & 23.75 & 31.25 & \underline{30.75} \\
ReasonRank & 33.50 & \underline{27.25} & 34.00 & 26.00 \\
Long-context LLM & 33.50 & 25.50 & 34.00 & 27.25 \\
\rowcolor[RGB]{235,245,250} MemSifter & \textbf{38.50} & \textbf{28.00} & \textbf{40.50} & \textbf{31.25} \\
\bottomrule
\end{tabular}
}
\end{table}

\section{Case Study}
\label{sec:appendix-case-study}

To provide a qualitative perspective on MemSifter's effectiveness, we present case studies on LoCoMo~(Figure~\ref{fig:locomo}) and WebDancer~(Figure~\ref{fig:webdancer}). These examples show the memory proxy's reasoning process and the ranked sessions it selects for the working LLM.

\section{Deep Research Dataset Construction}
\label{sec:appendix-deep-research}

To rigorously evaluate the capability of memory systems in handling complex, multi-step information seeking tasks, we constructed a specialized ``Deep Research'' benchmark. The core search trajectories and reasoning traces are sampled from the \textbf{MiroVerse} dataset~\footnote{\url{https://huggingface.co/datasets/miromind-ai/MiroVerse-v0.1}}.

However, to differentiate the performance of advanced memory methods from standard baselines, we significantly increased the difficulty of the original data through two key modifications:

\textbf{(1) Noise Injection via Semantic Distractors:} We augmented the original search trajectories by injecting \textit{approximate search results}. These distractors are semantically related to the query but contain factually irrelevant or misleading details. This design forces the memory model to perform precise discrimination based on logical consistency rather than simple keyword matching.

\textbf{(2) Context Extension:} We concatenated multiple search iterations and intermediate reasoning steps to create significantly longer context windows. 

This constructed environment serves as a \textbf{stress test} for retrieval-augmented generation. It evaluates whether a method can effectively sift through massive volumes of noisy search logs and reasoning chains to extract the specific needle-in-a-haystack evidence required for a correct final answer.

\begin{figure*}[p]
\begin{mdframed}[backgroundcolor=gray!5, roundcorner=3pt, innerleftmargin=10pt, innerrightmargin=10pt, innertopmargin=5pt, innerbottommargin=5pt, nobreak=true]

\small
\sloppy

\textbf{Question:} Where am I planning to stay for my birthday trip to Hawaii?

\textbf{Answer: } Oahu.

\textbf{Reasoning:} To determine the most relevant sessions based on the two inputs provided (the Historical Interaction Information and the current chat context), I need to analyze both of them.

The current chat context that was specified is: "Where am I planning to stay for my birthday trip to Hawaii?"

Now, let's look at the Historical Interaction Information to find the most relevant sessions related to Hawaii.

From the History:

- <session0> talks about planning a birthday trip to Hawaii with the user asking about hiking trails on Kauai

- <session27> discusses a birthday trip to Hawaii in October, with the user asking about snorkeling spots on the island

- <session24> mentions planning a family vacation to Hawaii with information about resorts that can accommodate a large group

- <session27> again talks about the birthday trip with more details about snorkeling spots

- <session0> has more details about hiking trails on Kauai

- <session4> talks about a family reunion (with the user mentioning their grandfather's 80th birthday celebration)

- <session6> discusses planning a family reunion for next summer, with the user mentioning their nephew's high school graduation

- <session4> mentions the grandfather's favorite dishes

- <session2> has more details about the family reunion

- <session1> talks about the location of the family reunion (the cabin in the mountains)

I could also consider more:

- <session8> (more details about the family reunion)

- <session9> (more information about the activities)

- <session10> (more details about the trip)

- <session11> (more information about the location)

- <session12> (more details about the activities)

- <session13> (more information about the culture)

- <session14> (more details about the trip)

- <session15> (more information about the location)

- <session16> (more details about the activities)

- <session17> (more information about the culture)

Based on the topic consistency (all related to Hawaii), user need continuity (the user mentions planning a birthday trip), and detail overlap (the sessions have information that matches the current context), the most relevant sessions are:

$\cdots$ \textit{(Truncated Verification Process)} $\cdots$

I think I can rank these in descending order of relevance as follows:

1. <session0> (high relevance - talks about hiking trails, with the user mentioning the birthday trip)

2. <session27> (more details about the trip, with the user mentioning snorkeling spots)

3. <session24> (more information about the family vacation with details about the accommodations)

4. <session4> (more details about the family reunion with the user mentioning the grandfather's birthday)

5. <session6> (more information about the family reunion with the user mentioning the nephew's graduation)

6. <session1> (more details about the location and activities)

7. <session2> (more information about the food)

8. <session3> (more details about the history)

9. <session5> (more information about the trip)

10. <session7> (more details about the culture)

This ranking is based on the topic consistency (all related to Hawaii), user need continuity (the user mentions planning a birthday trip), and detail overlap (the sessions have information that matches the current chat context).
</think>

<ranking>0,27,24,4,6,1,2,3,5,7</ranking>

\end{mdframed}
\caption{Case study for the reasoning trajectory on LoCoMo.}
\Description{A framed case study showing the question, answer, proxy reasoning trajectory, and ranked memory sessions for a LoCoMo example.}
\label{fig:locomo}
\end{figure*}

\begin{figure*}[p]
\begin{mdframed}[backgroundcolor=gray!5, roundcorner=3pt, innerleftmargin=10pt, innerrightmargin=10pt, innertopmargin=5pt, innerbottommargin=5pt, nobreak=true]

\small
\sloppy

\textbf{Question:} Which actor portrayed the father of the main character in the adaptations of a popular book series by J.K. Rowling, known for playing the role of the main character's father in a widely recognized film series?

\textbf{Answer: } Adrian Rawlins.

\textbf{Reasoning:} Let me analyze the two inputs provided: the Historical Interaction Information and the current chat context, to rank the most relevant sessions.

The current chat context given is: "Which actor portrayed the father of the main character in the adaptations of a popular book series by J.K. Rowling, known for playing the role of the main character's father in a widely recognized film series?"

Now, let's look at the Historical Interaction Information to find the most relevant sessions related to J.K. Rowling's work.

From the list, I can see:

- <session63>: Which actor portrayed the father of the main character in the adaptations of a popular book series by J.K. Rowling, known for playing the role of the main character's father in a widely recognized film series?

- <session9>: Identifying the Son of Paul Simon (with information about Paul Simon, the performer of "Graceland", and his sons)

- <session33>: Who Played Harry Potter's Father in the Film Adaptations (with information about Adrian Rawlins, who played James Potter, Harry Potter's father)

- <session21>: Final Response: Country of Origin for the Album '69 Love Songs' (with information about the album and its origin)

- <session2>: About the TV show 'Gotham' (with information about its production and location)

- <session0>: About a High Court jurist (with information about their work and background)

- <session1>: About the economic outlook of India (with data from Deloitte)

- <session2> (again, about the TV show 'Gotham')

- <session3>: About the financial regulations of a university (with information about the implementation date)

- <session4>: About the ownership structure of a company (with details about its structure)

- <session5>: About the history of a region (with information about its division and the population)

- <session6>: About a sports celebrity dinner (with details about the event)

- <session7>: About a lyricist born in the Palatinate region (with information about their work and background)

- <session8>: About the UNESCO World Heritage Sites (with information about the sites and their status)

- <session9>: More information about Paul Simon

- <session10>: About the Diocese of Bath and Wells (with details about its structure and the bishop)

- <session11>: About the percentage composition of a compound

- <session12>: About the seat of Brown County, Wisconsin

- <session13>: About the Court of the Lord Lyon (with information about a ruling)

- <session14>: About Oriel College Oxford (with details about its history and the alumni)

- <session15>: About the Visalia Fox Theatre (with information about its history and connection to a musician)

- <session16>: About the ethical mining of rose quartz (with details about the industry)

- <session17>: About a medical trial (with information about the research and methodology)

- <session18>: About the Hudson Bay drainage basin (with details about its geography and history)

- <session20>: About an infomercial-focused television channel in Australia

- <session21> (again, about the album '69 Love Songs')

- <session22>: About the IETF 122 event (with details about the conference)

- <session23>: About the van der Waals equation (with mathematical information)

- <session24>: About the South Atlantic (with information about its geography and history)

- <session25>: About the Indonesian Film Festival (with details about its history and significance)

$\cdots$ \textit{(Truncated Verification Process)} $\cdots$

Based on the topic consistency (the most relevant sessions should match the current chat topic) and user need continuity (they are related to the user's current need), the most relevant sessions for the current chat context are:

$\cdots$ \textit{(Truncated Verification Process)} $\cdots$

After reviewing, the most relevant sessions (highest relevance) are:

1. <session63> (the user mentioned the topic in the current chat context)

2. <session33> (more details about the same topic)

3. <session9> (more information about the author and his family)

4. <session21> (more details about the work)

5. <session2> (more information about the production)

6. <session0> (more details about the person)

7. <session1> (more information about the work)

8. <session3> (more details about the implementation)

9. <session4> (more information about the company)

10. <session5> (more details about the history)

This ranking is based on topic consistency (the sessions match the current chat topic) and user need continuity (they are related to the user's current need for information about the actor who portrayed the father of the main character in the adaptations of J.K. Rowling's book series).
</think>

<ranking>63,33,9,21,2,0,1,3,4,5</ranking>

\end{mdframed}
\caption{Case study for the reasoning trajectory on WebDancer.}
\Description{A framed case study showing the question, answer, proxy reasoning trajectory, and ranked memory sessions for a WebDancer example.}
\label{fig:webdancer}
\end{figure*}

\end{document}